\documentclass[5p]{elsarticle}

\usepackage{ulem} 
\usepackage{hyperref}
\usepackage{rotating}
\usepackage{multicol}
\usepackage{lipsum}
\usepackage[english]{babel}
\usepackage[section]{placeins}
\usepackage{caption}
\usepackage{subcaption}
\usepackage[labelfont=bf,textfont=md]{caption}
\usepackage{dashrule}
\usepackage[binary-units=true]{siunitx}
\usepackage{indentfirst}
\usepackage{graphicx}
\usepackage{array}
\usepackage{wrapfig}
\usepackage{multirow}
\usepackage{tabu}
\usepackage{algorithmic}
\usepackage{graphicx,url}
\usepackage{microtype}
\usepackage{float}
\usepackage{makecell}
\usepackage{units}
\usepackage{mathtools}
\usepackage{booktabs}
\usepackage{algorithm,algorithmic}
\usepackage{verbatim}
\usepackage{setspace}
\usepackage{siunitx}
\usepackage{amssymb}
\usepackage{amsmath} 
\usepackage{numprint}
\usepackage{colortbl}%
\usepackage{nomencl}
\usepackage[acronym,nogroupskip,nonumberlist,nopostdot]{glossaries}
\newcommand{\La}{\mathcal{L}}

\newacronym{qos}{QoS}{Quality of Service}
\newacronym{5g}{5G}{5th Generation of Wireless Systems}
\newacronym{uhd}{UHD}{Ultra-high-definition video}
\newacronym{qoe}{QoE}{Quality of Experience}
\newacronym{ilp}{ILP}{Integer Linear Programming}
\newacronym{smart-fl}{SMART-FL}{Multimedia  Service Placement Problem based on Facility Location}
\newacronym{tiptop}{TIPTOP}{mul\textbf{T}imedia serv\textbf{I}ce \textbf{P}lacemen\textbf{T} problem based on facility l\textbf{O}cation aware of data traffic \textbf{P}rediction}
\newacronym{mmpp}{MMPP}{Multimedia Microservices Placement Problem}
\newacronym{fc}{FG}{Fog Computing}
\newacronym{cc}{CC}{Cloud Computing}
\newacronym{ec}{EC}{Edge Computing}
\newacronym{iot}{IoT}{Internet of Things}
\newacronym{tcs}{TCS}{Traffic Control System}
\newacronym{flp}{FLP}{Facility Location Problem}
\newacronym{cflp}{CFLP}{Capacitated Facility Location Problem}
\newacronym{uflp}{UFLP}{Uncapacitated Facility Location Problem}
\newacronym{ran}{RAN}{RadioAccess Network}
\newacronym{fran}{FRAN}{Fog Radio Access Networks}
\newacronym{vod}{VoD}{Video on Demand}
\newacronym{cdr}{CDR}{Call Detail Record}
\newacronym{mips}{MIPS}{Millions of Instructions Per Second}
\newacronym{dbscan}{DBSCAN}{Density Based Spatial Clustering of Application with Noise}
\newacronym{cfg}{CFC}{Cloud-Fog Computing}
\newacronym{lstm}{LSTM}{Long Short-Term Memory}
\newacronym{arima}{ARIMA}{Autoregressive Integrated Moving Average}
\newacronym{c-ran}{C-RAN}{Cloud Radio Access Network}
\newacronym{npp}{NPP}{Network Planning Problem}
\newacronym{mae}{MAE}{Mean Absolute Error}
\newacronym{rmse}{RMSE}{Root Mean Square Error}
\newacronym{kf}{KF}{Kalman Filter}
\newacronym{rnn}{RNN}{Recurrent Neural Networks}
\newacronym{fl}{FL}{Fuzzy logic}
\newacronym{arima-pred}{ARIMA-PRED}{Autoregressive Integrated Moving-prediction}
\newacronym{lstm-pred}{LSTM-PRED}{Long Short-Term Memory-prediction}
\newacronym{da}{DA}{Dynamic Adjustment}
\newacronym{qoeap}{QoE-AP}{QoE-Aware Placement}
\newacronym{bps}{BPS}{Bits Per Second}
\newacronym{gru}{GRU}{Gated Recurrent Unit}

\graphicspath{ {COMCOM/} }

\begin{document} 

\begin{frontmatter}
\title{Multimedia Services Placement Algorithm for Cloud-Fog Hierarchical Environments}
\author[UNICAMP]{Fillipe Santos}
\ead{fillipe@lrc.ic.unicamp.br}
\author[UFRN]{Roger Immich}
\ead{roger@imd.ufrn.br}
\author[UNICAMP]{Edmundo R. M. Madeira}
\ead{edmundo@ic.unicamp.br}

\cortext[cor]{Corresponding author}

\address[UNICAMP]{University of Campinas.
Campinas, SP, Brazil}
\address[UFRN]{Federal University of Rio Grande do Norte
Natal, Brazil}


\begin{abstract}
With the rapid development of mobile communication, multimedia services have experienced explosive growth in the last few years.
The high quantity of mobile users, both consuming and producing these services to and from the \gls{cc}, can outpace the available bandwidth capacity.
\gls{fc} presents itself as a solution to improve on this and other issues.
With a reduction in network latency, real-time applications benefit from improved response time and greater overall user experience.
Taking this into account, the main goal of this work is threefold. 
Firstly, it is proposed a method to build an environment based on \gls{cfg}. 
Secondly, it is designed two models based on \gls{arima} and \gls{lstm}.
The goal is to predict demand and reserve the nodes’ storage capacity to improve the positioning of multimedia services.
Later, an algorithm for the multimedia service placement problem which is aware of data traffic prediction is proposed.
The goal is to select the minimum number of nodes, considering their hardware capacities for providing multimedia services in such a way that the latency for servicing all the demands is minimized.
An evaluation with actual data showed that the proposed algorithm selects the nodes closer to the user to meet their demands.
This improves the services delivered to end-users and enhances the deployed network to mitigate provider costs.
Moreover, reduce the demand to Cloud allowing turning off servers in the data center not to waste energy.
\end{abstract}

\begin{keyword}
\texttt Cloud-to-Fog Networks \sep Multimedia Services \sep Placement Strategies
\end{keyword}

\end{frontmatter}

\section{Introduction}
\label{introduction}
In recent years there has been a rapid proliferation of a wide range of real-time multimedia services, such as video on demand, video conferencing, broadcast of interactive 3D environments, high definition videos, streaming video with 4k/8k resolution \gls{uhd}, among others~\cite{Immich2019}.
These services already account for the majority of global traffic and by 2021/2022 will flood mobile networks requiring unprecedented high speed and low latency~\cite{batalla2017efficient, Pisani2020}.

According to technical reports provided by Cisco Systems~\cite{forecast2019cisco}, 73\% of all global IP traffic generated on the Internet in 2019 was related to video traffic over IP and 1\% related to gaming traffic, with projections that these percentages will be 82\% and 4\%, respectively, for the years 2021/2022.
In fact, the adoption of the \gls{5g} will allow this growth to be even greater due to its high bandwidth capacity and low latency.
Also, the most recent studies on consumer habits during the pandemic, and what may remain later, show that multimedia service traffic peaked at $\approx$ 60\% higher than January levels when the lockdowns started in some countries~\cite{lutu2020characterization, favale2020campus}.

Notwithstanding the numerous advantages that \gls{cc} offers, such as scalability, security, and flexibility, multimedia services require constant, continuous flow of packets with low latency~\cite{Immich2015a}, requirements that it occasionally does not provide~\cite{zhu2011multimedia}.

Adapting these services in this environment is a non-trivial task~\cite{bittencourt2018}.
By the way, even with the improvements in wireless technologies offered by the \gls{5g}, reliable, high-quality video delivery still poses several challenges, such as dealing with a large number of heterogeneous devices and meeting the increasing requirements of users.
To overcome these and other problems, it is desirable to use a distributed architecture that stores and processes services logically between the Cloud and the data source~\cite{cheng2020edge}.

\gls{fc} and \gls{ec} present themselves as a joint solution to meet these latency-sensitive services, where the management of all resources occurs in a coordinated and tiered manner, from the Cloud to the end devices.
The main goal is to allocate Cloud resources physically closer to end users~\cite{taleb2014lightweight}. 
These architectures share similar benefits compared to \gls{cc}, including a reduction in latency to milliseconds, decreased network congestion, and real-time recognition of users' geographic location~\cite{mahmud2018fog}. 

Nodes are hierarchically organized in tiers in these environments, from the Cloud to the end-users.
They are also namely fog or edge nodes.
These nodes are infrastructures that can provide resources for services that can be executed in a distributed and independent way as microservices, available closer to end-users~\cite{solutions2015unleash}.
On the one hand, nodes belonging to the same tiers have similar networks (latency, download, and upload) and computational (storage and processing) resources.
On the other hand, nodes of different tiers have distinct features.
The non-availability of these environments for simulation makes the evaluation of these algorithms a challenge.
Also, the hierarchical, distributed, and heterogeneous nature of computational instances makes the positioning of the multimedia services in these environments a challenging task~\cite{rosario2018service,Gama2018}. 

Placing multimedia services in these environments can be related to the \gls{flp}~\cite{c2019location}.
In short, this problem refers to optimal placement of facilities (resources) that an organization has to meet the clients' requirements so while considering a set of constraints like the distance between resources and clients or competitors' facilities~\cite{farahani2009facility}.  
In this problem, the facilities may or may not have capacity constraints for storage, which categorize the problems into capacitated and uncapacitated variants~\cite{anas1982residential}.
The \gls{cflp} is the basis for many practical applications, where the facilities have a limit on the number of customers it can serve.
For an \gls{uflp} however, the assumption made is that an arbitrary number of customers can be connected to a facility.

To further improve the service delivery, algorithms based on traditional and deep learning models, such as \gls{arima} and \gls{lstm}, can be adopted to extract features from the telecommunication activity dataset and find correlations among them.
The goal is to predict future demand and reserve the nodes' storage capacity to improve the positioning of multimedia services.
The \gls{arima} model is used to understand time series or predict a point in the future.
Otherwise, \gls{lstm} model is a special type of \gls{rnn} capable of learning long-term dependencies~\cite{zhang2003time}.

In previous work~\cite{mypaper}, we have introduced a novel algorithm to the \gls{mmpp} modeled as \gls{cflp}. 
We extend this algorithm by considering the network traffic prediction in this work.
All contributions are summarized below:
\begin{itemize}
\item The design and implementation of a new method to build an environment based on \gls{cfg}.
Nodes are organized hierarchically in tiers in this environment, from Fog to Cloud.

\item It is designed two models based on \gls{arima} and \gls{lstm} to solve the traffic forecasting problem.
The goal is to predict demand and reserve the nodes' storage capacity to improve the positioning of multimedia services.

\item An algorithm for the \gls{tiptop}.
The goals are to select the minimum number of nodes, considering their hardware capacities for providing multimedia services in such a way that the latency for servicing all the demands is minimized.
\end{itemize}

The performance assessment was carried out in MultiTierFogSim using two months of real-world mobile network traffic data in Milan, Italy.
First, the proposed algorithm is evaluated considering six snapshots selected in particular days and hours based on the traffic intensity to assess the performance of the \gls{smart-fl} algorithm in different circumstances.
Later, the performance assessment is based on predicted mobile traffic one month.
In this case, the results are compared considering four strategies to place multimedia services and are evaluated in terms of latency, package delivery, attempted requests, and network usage.

The results show that the proposed algorithm can balance the fog nodes' geographical location, hardware capacity, and the users' location.
The positioning becomes more efficient through data traffic prediction due to previously reserved nodes' storage.
Furthermore, using the information obtained in this work, it is possible to implement a strategy for shutting down servers in the Cloud to save energy.
It is worth mentioning that, the proposed scheme can be adapted for other services, e.g., \gls{tcs}, \gls{iot} applications, augmented reality, and others that require \gls{cc}, \gls{fc}, and \gls{ec} as well as service migration.

The remainder of this work is organized as follows.
Section~\ref{sec:realtedWork} gives an overview of the main related work. 
Section~\ref{sec:Cloud-to-Fog-Networks} presents the design of Cloud-Fog hierarchical environments.  
Section~\ref{sec:the-design-of-smart-fl} presents the formulation for the multimedia services placement problem.
Section~\ref{sec:traffic-forecast} describes how the prediction models are analyzed and implemented.
Section~\ref{sec:evaluations-and-experiments} details the experimental method and results.
Finally, Section~\ref{sec:conclusion} presents the conclusion and future work.

\section{Related Work}
\label{sec:realtedWork}

There are several proposals in the literature to optimize multimedia delivery over hierarchical networks. This section is divided into two parts to cover all aspects of this work, namely Cloud-Fog hierarchical environments~(\ref{sec:trab-rela-ambientes-hierarquico}) Multimedia services placement in Cloud-Fog hierarchical environments~(\ref{sec:trab-rela-multimedia-placement}).

\subsection{Cloud-Fog hierarchical environments}
\label{sec:trab-rela-ambientes-hierarquico}
Cloud-Fog hierarchical environments have also been proposed to optimize various issues related to this domain, from identifying the most appropriate grouping of base stations to share Cloud resources or even minimizing the distance between servers and access points throughout the city~\cite{chen2017complementary, barlacchi2015multi}.

A proposal for positioning fog nodes to reduce the costs associated with their deployment and maintenance considering variable demands in time is proposed in~\cite{c2019location}. 
The set of selected nodes compose the hierarchical environment.
Based on actual data, the results show that there is an improvement in end-user service that can be achieved in conjunction with minimizing costs by deploying a smaller number of servers in the infrastructure. 
Besides, costs can be reduced further if a limited blocking of requests is tolerated.
However, like previous work, the type of workload is simulated, and the number of tiers and nodes are limited by one and three, respectively.

A solution to the base station grouping problem for sharing \gls{c-ran} resources is proposed in~\cite{chen2017complementary}. 
The solution aims to group neighboring base stations with complimentary traffic patterns so that the traffic volume processed in \gls{c-ran} is balanced, requiring fewer resources. 
The results show that this collation scheme reduces deployment cost by 12,88\%.
The data set used was made available by Telecom Italia~\cite{barlacchi2015multi}.
However, the number of tiers and nodes are limited by two and four, respectively.
In our proposed method, the number of tiers varies.

A framework for partitioning a set of base stations into groups and processing the data in a shared data center is proposed in~\cite{bhaumik2012cloudiq}. 
This partitioning and scheduling framework saves up to 19\% of computing resources for a one in 100 million probability of failure.
However, the adoption of only one data center can result in delays between the distant base stations and the data center.
Also, the type of workload is unrealistic, and the number of tiers and nodes are limited by two and five, respectively.
Further, it is not considered balanced workload and location preference.
In contrast, we experiment with publicly available real-world mobile traffic data set and, as mentioned, the number of tiers varies, taking into account the balanced workload and location preference.

The placement of edge servers considering capacity constraints is modeled as a Capacitated Location-Allocation problem by~\cite{lahderanta2019edge}.
The goal is to minimize the distance between servers and access points throughout the city.
The results show that the proposed algorithm can provide optimal solutions that minimize distances and offer a balanced workload with sharing according to node capacity constraints.
As in previous work, they used an unrealistic data set and the number of tiers is limited to seven, without balanced workload and location preference.

A proposal for locating fog nodes with limited battery support for mobile users capable of processing high demands with low latency restrictions is proposed by~\cite{9062311}.
The conclusion is that the heuristic solution produces accurate results when compared to data generated by the \gls{ilp}, thus allowing significant energy savings for end-users.
However,the number of tiers is limited to seven, without balanced workload and location preference.

In this work, a method is proposed to creating Cloud-Fog hierarchical environments.
The method uses a bottom-up approach, starting from a set $BS = \{bs_ {1}, bs_ {2}, ..., bs_ {bs} \} $ of base stations and organizing new nodes hierarchically into tiers, producing a hierarchical Cloud-Fog environment.
Fog nodes are grouped to facilitate resource pooling and work collaboratively, reducing effort in obtaining an optimal node towards service deployment.
These groups of fog nodes are also linked to enable service migration whenever necessary.

Table~\ref{table:cloud-fog-environments-works} lists the related works and classifies their contributions with respect to five characteristics.
The first column represents the related works.
The second column represents the type of workload.
The third column refers to the number of tiers.
The fourth column shows the number of nodes able to services.
The fifth column refers to if the work considers balanced workload.
Finally, the sixth column indicates if the work considers location preference.
 
\begingroup
\renewcommand\arraystretch{1.5}
\begin{table}[!htb]
\centering
\caption{Comparison table of related works.}
\scalebox{0.88}{
\begin{tabular}{@{}cccccc@{}}
\toprule
\thead{\rotatebox[origin=c]{80}{\textbf{Research work}}} & \thead{\rotatebox[origin=c]{80}{\textbf{Workload}}} &  \thead{\rotatebox[origin=c]{80}{\textbf{Number of tiers}}} & \thead{\rotatebox[origin=c]{80}{\textbf{Number of nodes}}} & \thead{\rotatebox[origin=c]{80}{\textbf{Balanced workload}}} & \thead{\rotatebox[origin=c]{80}{\textbf{Location preference}}} \\ 
\midrule 
C da Silva el at. \cite{c2019location} & Simulated  & 1 & 3 & \checkmark & \checkmark \\ 
Chen el at. \cite{chen2017complementary} & Phone call & 2 & 4 & \checkmark & \checkmark \\ 
Bhaumik el at. \cite{bhaumik2012cloudiq} & Simulated  & 2 & 5 & $\times$ &  $\times$ \\ 
L{\"a}hderanta el at. \cite{lahderanta2019edge} & Simulated  & 7 & 4 & $\times$ & $\times$  \\
Silva el at. \cite{9062311} &  Phone call  & 7 & N & $\times$ & $\times$ \\
\textbf{Proposed solution} & Phone call & N & N & \checkmark & \checkmark \\
\bottomrule
\end{tabular}}
\label{table:cloud-fog-environments-works}
\end{table}
\endgroup

\subsection {Multimedia services placement in Cloud-Fog hierarchical environments}
\label{sec:trab-rela-multimedia-placement}

Many research efforts have been conducted to address the issue of reducing latency to deliver multimedia services in the context of \gls{cc}, \gls{fc}, and \gls{ec}.
They range from changes in the Cloud architecture to the use of fog/edge nodes to reduce network delay and improve \gls{qoe}~\cite{shi2021hierarchical, kharel2018multimedia}.

Table~\ref{table:related-works} lists the related works and classifies their contributions with respect to five characteristics.
The first column represents the related works.
The second column refers to if the work was performed in a Cloud-Fog hierarchical environment.
The third column shows the number of nodes able to deploy multimedia services.
The fourth column refers to if the work considers the nodes' hardware capacity.
The fifth column indicates if the work considers multiple requests for multimedia services.
Finally, the sixth column indicates if the work considers network traffic prediction.
The symbol $'?'$ means that the authors did not include the information.

\begingroup
\renewcommand\arraystretch{1.5}
\begin{table*}[t]
\centering
\caption{Comparison table of related works.}
\scalebox{0.9}{
\begin{tabular}{@{}cccccc@{}}
\toprule
\textbf{\thead{Research work}} & \textbf{\thead{Cloud-Fog hierarchical\\environments}} & \textbf{\thead{Number of\\nodes}} & \textbf{\thead{Node\\capacity}} & \textbf{\thead{Multiple\\requests}} & \textbf{\thead{Network traffic\\prediction}}\\ 
\midrule 
Souza el at. \cite{souza2016handling} & Cloud/Fog  & 7 & \checkmark & $\times$ & $\times$ \\
Fang Shi al. \cite{shi2021hierarchical} & Cloud/Fog & ? & \checkmark & \checkmark & $\times$ \\
Kharel et al. \cite{kharel2018multimedia} & Cloud/Fog & 84 & \checkmark & $\times$ & $\times$ \\
Kryftis et al. \cite{kryftis2017resource} & Cloud  & ? & $\times$ & $\times$ & \checkmark \\
Mahmud et al. \cite{mahmud2019quality} & Fog & 4-10 & \checkmark & \checkmark & $\times$ \\
Sai et al. \cite{sai2020cooperative} & Cloud/Fog  & 4 & \checkmark & \checkmark & $\times$ \\
\textbf{Proposed solution} & Cloud/Fog & 1160 & \checkmark & \checkmark & \checkmark \\
\bottomrule
\end{tabular}}
\label{table:related-works}
\end{table*}
\endgroup


A \gls{qos}-aware service allocation problem for Cloud-Fog architectures as an integer optimization problem was proposed in~\cite{souza2016handling}.
The technique combines the Cloud-Fog operations and can accomplish high system capacity, granting low latency for requested services.
The hierarchy of a layer is determined by capacity, vicinity, and reachability to end-users.
The authors concluded that service distribution benefits among multi-tier fog nodes because they avoid the high delay access on the cloud layer.
Nevertheless, the effect on time overhead created by the service distribution with a vast number of fog nodes for mobile users is not considered.
In normal conditions, this may be analyzed. 
However, in an unusual situation, this could be a problem.
For example, resources can be consumed by a large group of users.
The proposed algorithm has a time and space complexity of $\mathcal{O}(mn\log{\frac{n^2}{m}} + nlog(m))$ and $\mathcal{O}(n+m)$, respectively; where $n$ and $m$ are the services requiring set size and \gls{iot} nodes, respectively.

A hierarchical content delivery network in a randomly distributed interference environment was proposed by~\cite{shi2021hierarchical}.
The optimization problem is modeled as a combinatorial optimization problem.
The authors concluded that the proposed algorithms improve the users' cache hit probability and provide more flexible cooperative transmission opportunities.
However, a collaborative resource strategy in multi-tier fog nodes receives more attention.
Also, the user preference model was not validated.
The proposed algorithm has a time and space complexity of $\mathcal{O}(mn\log{\frac{n^2}{m}}log(m))$ and $\mathcal{O}(n+m)$, respectively; where $n$ and $m$ are the number of areas with service requests and nodes, respectively.

A hierarchical \gls{fc}-based multimedia streaming that reduces latency and minimizes internet bandwidth consumption for passengers traveling in any vehicle is proposed by~\cite{kharel2018multimedia}. 
The result acquired from the simulation showed that the secondary and primary fog stations located at the edge were considered the better and best-case scenario to allocate multimedia services regarding \gls{qos} and \gls{qoe}.
However, the proposed hierarchical \gls{fc} does not consider that the number of vehicles requesting multimedia services can be more significant in practice.
Also, they do not consider that the demand for multimedia services and nodes' hardware capacity vary over time.
The proposed algorithm has a time and space complexity of $\mathcal{O}(m(n+\log{n}))$ and $\mathcal{O}(n)$, respectively; where $n$ and $m$ are the number of vehicles and nodes, respectively.

A network architecture that utilizes novel resource prediction models for optimal selection of multimedia content provision methods is proposed by~\cite{kryftis2017resource}.
The authors also present two algorithms for the delivery of these contents.
The results show a reduction in congestion and an 80\% success rate of the services' transmission.
The prediction engine is accurate, and in general, the content delivery process benefits from using the prediction model and algorithms. 
However, services are offered in the Cloud, where, as already discussed, \gls{cc} does not provide a satisfactory \gls{qoe} in areas with high demand for these services. 
Besides, new prediction models have emerged, such as \gls{lstm} networks that may present greater accuracy in network traffic prediction than models analyzed by the authors.
The proposed algorithm has a time and space complexity of $\mathcal{O}(m+n\log{n})$ and $\mathcal{O}(n+m)$, respectively; where $n$ and $m$ are the number of clients and nodes, respectively.


A \gls{qoe}-aware application placement policy for distributed \gls{fc} environments is proposed by~\cite{mahmud2019quality}. 
The results indicate that the policy significantly improves data processing time, resource affordability, and service quality.
However, a solution for service placement considering a large volume of data in the decentralized architecture in \gls{fc} can cause network congestion~\cite{Fog-orchestration-for-the-Internet-of-Everything:state-of-the-art-and-research-challenges}.
The proposed algorithm has a time and space complexity of $\mathcal{O}(\log{\frac{B}{M} (m+n)(m+n\log{n})}) $ and $\mathcal{O}(B+m)$, respectively; where $B$, $m$, and $n$ are the number of clients, caches, and nodes, respectively.

An efficient collaborative content delivery and caching strategies in a 5G network
is proposed in\cite{sai2020cooperative}.
They propose a cache placement algorithm based on a greedy heuristic algorithm to solve this energy efficiency problem by optimizing cache placement.
The authors conclude that the proposed system reduces the number of interference and improves the system throughput.
However, implementing intelligent mechanisms that assess network and node conditions for dynamical deployment of these services could be more appropriate~\cite{osanaiye2017cloud}.
The proposed algorithm has a time and space complexity of $\mathcal{O}(m\log{n} + n\log{n}) $ and $\mathcal{O}(n+m)$, respectively; where $m$ and $n$ are the number of caches and clusters, respectively.

This work proposes an algorithm to select the minimum number of nodes for multimedia services placement aware of traffic prediction.
It is taken into consideration their hardware capacities and network traffic predicted for providing multimedia services in such a way that the latency for servicing all the demands is minimized.
The proposed algorithm has a time and space complexity of $\mathcal{O}(m\log{m})$ and $\mathcal{O}(m)$, respectively; where $m = n_c \times n_f$, where $n_c$ and $n_f$ are the number of regions with multimedia services requests and nodes, respectively~\cite{shu2013fast}.
In this work, regions are grids of 235 $\times$ 235 $m^2$.

\section{The design of Cloud-Fog Hierarchical Environments}
\label{sec:Cloud-to-Fog-Networks}
The fast-growing mobile network data traffic poses great challenges for operators to increase their data processing capacity in base stations efficiently. With the Cloud Radio Access Network~(Cloud-RAN) and \gls{fc}, the data processing units can now be centralized in a data center and shared among several base stations. 
Also, clustering base stations can reduce the deployment cost and energy consumption with complementary traffic patterns to the same data center. 
To evaluate the proposed algorithm and contribute to this challenge, initially it is proposed a method to build an environment based on Cloud-Fog Computing. 

This environment is based on the Cloud-Fog architecture, which provides a virtualized, hierarchically organized, distributed computing platform. All nodes are a small facility that hosts dedicated servers capable of processing end-user workload.
The nodes closer to end-users are expected to have a smaller capacity, increasing towards the Cloud in the infrastructure, forming a Cloud-Fog Hierarchical Environment in which any device can access the Cloud.

The proposed method uses a bottom-up approach, starting from a set $BS = {bs_ {1}, bs_ {2}, ..., bs_ {bs}}$ of base stations, and arranges new nodes hierarchically, from Fog to Cloud.

\begin{figure*}[htp!]
    \centering
        \includegraphics[width=.8\linewidth]{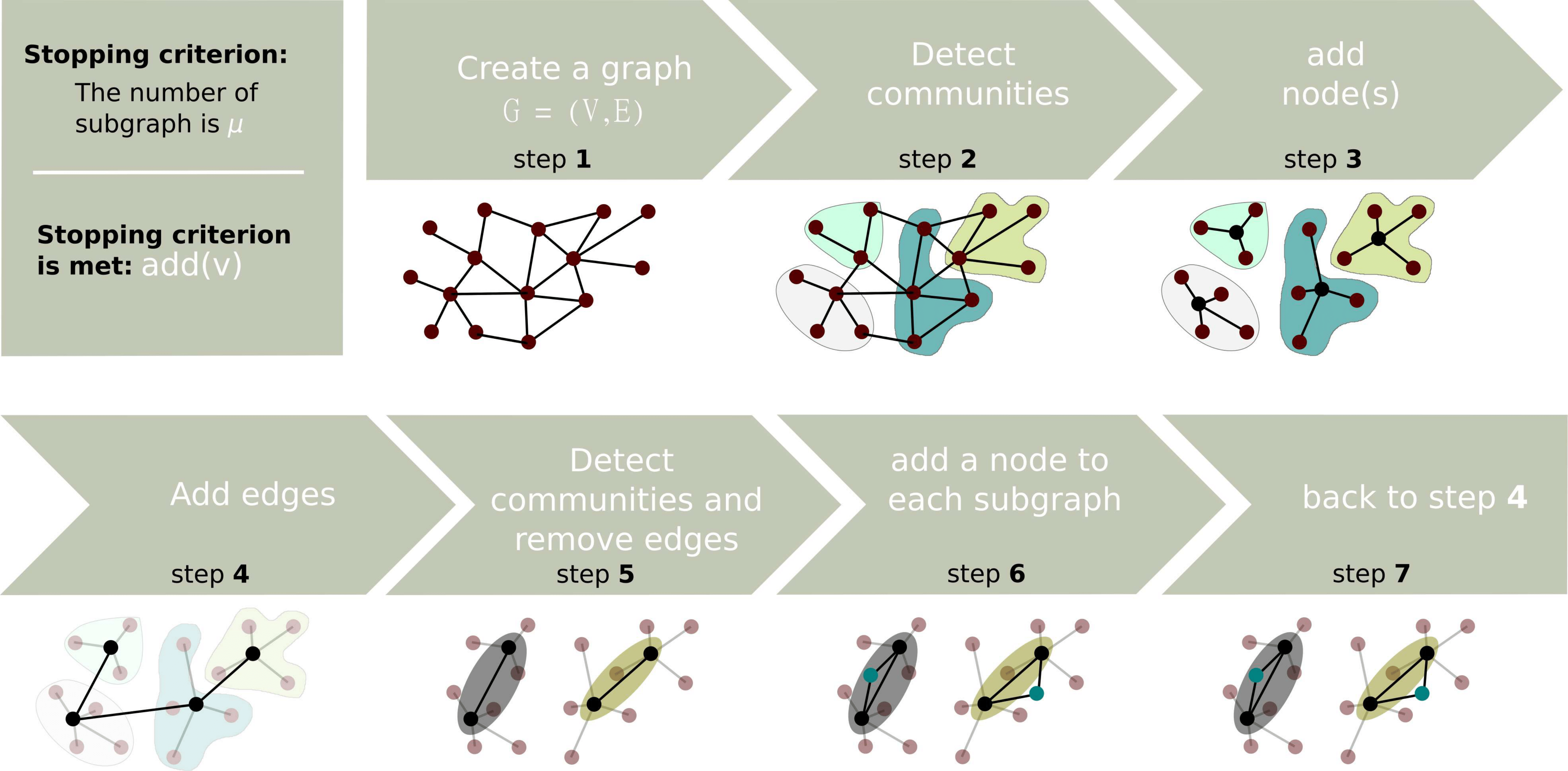}
    \caption{Method for design of Cloud-Fog Hierarchical Environments.}
\label{fig:method}
\end{figure*}

Initially, it is defines $\Xi$ as the connection between the base stations.
For $\Xi$, one can consider the distance $R$, signal strength, traffic similarity, etc.
Let $\textbf{G} = (V, \mathcal{E})$ be an unweighted undirected graph, where $V$ contains all base stations, i.e., $V \equiv BS$; 
and $\mathcal{E}$ is the set of edges that represent the communication among base stations defined by $\Xi$. Method~1 and Figure~\ref{fig:method} describe all the steps.

The stopping criterion is defined by $\mu$ as the number of subgraphs of the penultimate tier and must always be observed at steps two and five.
When it is met, an upper-tier node that connects to the lower-tier nodes is added when the stopping criterion is met.

Considering \textbf{step 1}, nodes from other providers can be added as a vertex in $\mathcal{G}$.
The nodes added (\textbf{steps 2-6}) can also be considered as service providers and can be a solution to the \gls{npp}~\cite{luna2010evolutionary}.
Also, their geographic locations are not limited to the area of the region considered. 
That is, these nodes may be positioned geographically in other locations.

\begin{table}[htp!]
\renewcommand\arraystretch{1.3}
\centering
\scalebox{0.9}{
\begin{tabular}{l}
\toprule
\midrule
\multicolumn{1}{c}{\thead{\textbf{Method 1:} Method to build a\\Cloud-Fog hierarchical environment.}}\\
\midrule
\textbf{Input:} $BS$, $\Xi$, $\mu$\\
\textbf{Output:} $\mathcal{G}$\\
\midrule 
\textbf{Stopping criterion:} The number of subgraph is $\mu$.\\
\textbf{Stopping criterion is met:} Add an upper-tier node that\\connects to the lower tier nodes.\\
\midrule
\textbf{1} - Define an undirected unweighted graph $\mathcal {G} = (BS, \mathcal {E})$.\\
\textbf{2} - Detect communities in $ \mathcal {G}$.\\
\textbf{3} - For each community, an upper-tier node is added in $ \mathcal{G} $,\\which communicates with all the base station nodes that the\\community belongs to.
This step ends with removing the\\edges between all base stations.\\
\textbf {4} - Add edges between all nodes in the current tier.\\
\textbf {5} - Detect communities and remove edges between nodes of\\different communities.\\
\textbf {6} - For each subgraph, add an upper-tier node with edges\\between the node and the subgraph.\\
\textbf {7} - Go back to step \textbf{4}.\\
\midrule 
\bottomrule
\label{alg:meotodo-proposto}
\end{tabular}}
\end{table}

Community structure, also called clusters, groupings, or communities, are groups of vertices that are very likely to share common properties or have similar roles in the graph~\cite{xie2011community}.
The communities detected in \textbf{step 2} and \textbf{5} are represented by set $S\;=\;\{s_{1},\;s_{2},\;...,\;s_{s}\}$ and can be detected by any strategy~\cite{blondel2008fast}.

The proposed environment can be simulated in any Cloud-Fog-Edge Simulator, where it benefits from several advantages related to Cloud and Fog, such as location awareness, analysis capability for processing, as well as service migrations to and from any layer.

\section{Multimedia services placement algorithm}
\label{sec:the-design-of-smart-fl}
First, the \gls{smart-fl} algorithm introduced in previous work~\cite{mypaper} is presented. 
Later, an extension of this algorithm by considering the network traffic prediction is presented.

Addressing the multimedia services placement problem in Cloud-Fog hierarchical environments involves considering some specificities and criteria when designing the deployment strategies.

Regarding resource constraints, we consider that finite capabilities limit the nodes in terms of CPU, RAM, storage, and bandwidth.
Let $\La = \{ell_{1}, ell_{2}, ell_{3}, ..., ell_{x}\}$ be the set of nodes capable of processing and storing multimedia services.
This includes all nodes with maximum capacity $c_{max(\ell)}$, where $\ell\;\in\;\La$. 
Let $W\;=\;{w_1,\;w_2,\;w_3,\;...,\;w_z}$ be the set of multimedia services; $U\;=\;{u_1,\;u_2,\;u_3,\;...,\;u_k}$ the regions with multimedia services requests; and $k = |U|$ the number of regions with multimedia services.
Then, $G_{t}\;=\;\{u_{1}^{w_{1}},\;u_{2}^{w_{2}},\;...,\;u_{k}^{w_{z}}\}$ is the set of multimedia services $w\;\in\;W$ requested at time $t$ in region $u_{k}$;

Network links are bound by constraints such as latency, which need to be satisfied when deploying multimedia services.
Variable $lat(\ell, u_{k}^{w_{z}},t)$ is the latency offered by $\ell$ to service $w_z$ in region $u_k$ at time $t$, where $u_{k}^{w_{z}}\;\in\;G_{t}$.
Variable $x({\ell,u_{k}^{w_{z}},t})\;\geq\;0$ represents the multimedia service $w$ requested in region $u_k$ processed by node $\ell$ at time $t$, where $u_{k}^{w_{z}}\;\subseteq\;G_{t}$.

In this way, when placing these services, we need to respect the resource requirements, i.e., ensure that the resources of the components deployed on the infrastructure nodes do not exceed their capabilities.

Therefore, a solution for the multimedia services placement problem in Cloud-Fog hierarchical environments is modeled as \gls{cflp}, namely \gls{smart-fl}, where
(\textbf{i}) nodes are the potential facility sites,
(\textbf{ii}) multimedia services requests are the demands,
(\textbf{iii}) nodes' storage capacity and the users' demand are part of the constraint set, and
(\textbf{iv}) multimedia services correspond to the type of service considered.
An integer-optimization model can be specified as follows:

\begin{align}
    \shortintertext{Minimize}   
    \MoveEqLeft \displaystyle\sum_{\ell \in \La} y({\ell, w_z, t}) + \sum_{\ell \in \La}\, \sum_{u_{k}^{w_{z}} \in G_{t}} lat(\ell, u_{k}^{w_{z}},t) \cdot x({\ell,u_{k}^{w_{z}},t})  \label{eq:facility-location-1}\\ 
    \shortintertext{subject to} 
    &\sum_{\ell \in \La} x({\ell,u_{k}^{w_{z}},t}) = u_{k}^{w_{z}} \;\;\;\;\;\;\;\;\;\;\;\;\;\;\;\;\;\;\;\;\;\;\;\;  \forall u_{k}^{w_{z}} \in G_{t} \label{eq:facility-location-2}\\
    &\displaystyle\sum_{g^{w} \in G_{t}} x({\ell,u_{k}^{w_{z}},t}) \leq c_{max}(\ell, t) \cdot y({\ell, w_{z}, t})\; \forall \ell \in \La \label{eq:facility-location-3}\\
    &x({\ell,u_{k}^{w_{z}},t}) \geq 0 \;\;\;\;\;\;\;\;\;\;\;\;\;\;\;\;\; \forall \ell \in \La \; and \; \forall u_{k}^{w_{z}} \in G_{t} \label{eq:facility-location-4}\\
    &y({\ell, w_{z}, t}) \in \{0,1\} \;\;\;\;\;\;\;\;\;\;\;\;\;\;\;\;\;\;\;\;\;\;\;\;\;\;\;\;\;\;\;\;\; \forall \ell \in \La& \label{eq:facility-location-5}\\
    &w_{z} \in W\label{eq:facility-location-6}\\
    &t \in [0, max\_simulation\_time]\label{eq:facility-location-7}
\end{align}

The objective function~\ref{eq:facility-location-1} is composed of two parts.
The first part selects nodes that minimize the associated costs.
The binary variable $y({\ell, w_z, t})\;=\;1$ indicates if multimedia service $w_z$ is deployed at node $\ell$ at time $t$, $y({\ell, w_z, t})\;=\;0$ otherwise.
The second part associates the latency and processing cost of the node $\ell$ to meet the multimedia service $w_z$ in region $u_k$ at time $t$.
The constraint in Equation~\ref{eq:facility-location-2} requires that the service $w_z$ requested in region $u_k$ processed by node $\ell$ at time $t$ must be satisfied.
The nodes' capacity is limited by the constraint in Equation~\ref{eq:facility-location-3}.
That is, if node $\ell$ is not activated, the demand satisfied by $\ell$ is zero. 
Otherwise, its capacity restriction is observed.
Finally, the constraints in Equations~\ref{eq:facility-location-4}-\ref{eq:facility-location-7} set the minimum values for the decision variables.
The \gls{ilp} model was coded using the Gurobi Optimizer solver~\cite{optimization2014inc}.
Gurobi is a commercial mathematical programming solver. 
t is possible to implement shared-memory parallelism, which efficiently exploits any number of processors and cores per processor. 
The solver uses an iterative process to converge on an optimal solution.


To place multimedia services aware of predicted mobile traffic, consider the performance of the \gls{smart-fl} algorithm. 
Let $t(n)$ be the current time.
Let $A_{t(n)}$ be the set of nodes selected to provide multimedia services at time $t(n)$ (performed by \gls{smart-fl} algorithm).
Let $P(A_{t(n+1)})$ be the set of nodes selected at $t(n)$ to provide multimedia services at $t(n+1)$, taking into account the traffic prediction at time $t(n)$ to $t(n+1)$.

Currently at $t(n+1)$, let $A_{t(n+1)}$ be the set of nodes selected at $t(n+1)$ to provide multimedia services at $t(n+1)$.
Thus, $Y\; = \; A_{t(n)} \; \cap \; P(A_{t(n+1)})$ contains the \textbf{reserved nodes} to provide multimedia services at $t(n+1)$, case $Y\;\neq \;\{\emptyset\}$.
Therefore, $\Gamma\; = \; Y \;\cup\; P(A_{t(n+1)})$ contains the \textbf{adequate nodes} to provide multimedia services at $t(n+1)$.

This means that the resources available by the nodes that provide multimedia services and that will be concluded at $t(n)$ belonging to $Y$ set are reserved to provide multimedia services at $t(n+1)$.
Thus, multimedia services that will be provided in $Y$, requested at $t(n+1)$, will not compete for resources with concurrent services at $t(n+1)$.
The advantage is that nodes in $\Gamma$ offer lower latency than nodes in $A_{t(n+1)}$.
The \gls{smart-fl} algorithm multimedia aware of predicted mobile traffic is called ~\gls{tiptop}.

\begin{figure}[htp!]
    \centering
        \includegraphics[width=.7\linewidth]{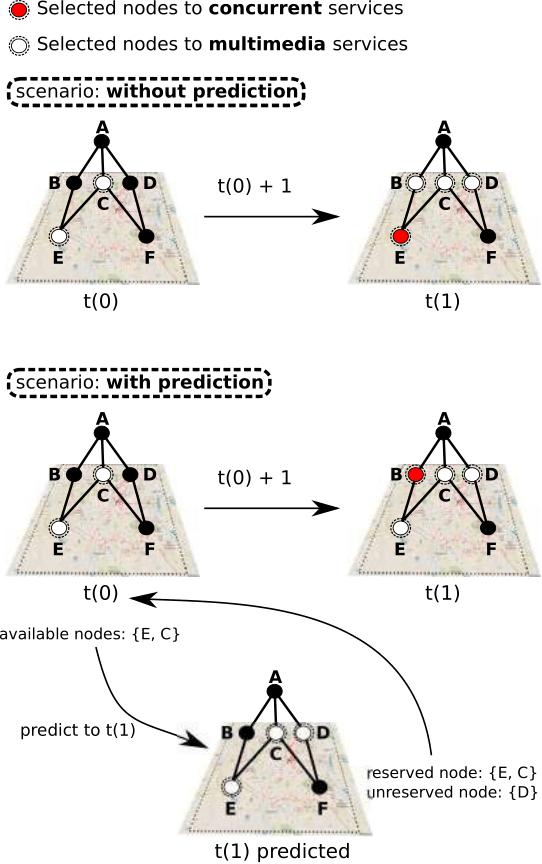}
    \caption{Multimedia services placement (scenario without and with mobile traffic prediction).}
\label{fig:previsao_facilitylocation}
\end{figure}

Figure~\ref{fig:previsao_facilitylocation} illustrates an example considering two scenarios: without and with mobile traffic prediction.
For the scenario without mobile traffic prediction, the nodes selected at $t(0)$ are $A_{t(0)} \; = \; \{E, C\}$.
Otherwise, the nodes selected at $t(1)$ are $A_{t(1)} \; = \; \{B, C, D\}$.
In this scenario, there are no reserved nodes to $t(1)$, and other services (i.e., services offered in \gls{tcs} and \gls{iot}), requested at $t(1)$, are provided by $E$ node.

For the scenario with mobile traffic prediction, the selected nodes are $A_{t(0)} \; = \; \{E, C\}$, $A_{t(1)} \; = \; \{B, C, D\}$, and $P(A_{1}) \; = \; \{E, C, D\}$ for $t(0)$, $t(1)$ and $t(1)$ (predicted), respectively.
In this way, $Y\; = \; \{B, C, D\} \; \cap \; \{E, C, D\} = \{C,D\}$;
 $\Gamma\; = \; Y \;\cup\; P(A_{t(n+1)}) = \{C,D\} \;\cup\; \{E, C, D\}$. 
Therefore, $Gamma \; = \; \{E, C, D\}$.

Considering that multimedia services provided by $A_{t(0)}$ will be concluded at $t(0)$ and $Y = \{E, C\}$ contains the reserved nodes to provide multimedia services at $t(1)$, the concurrent service that would be provided by node $E$ at $t(1)$ (scenario without mobile traffic prediction) is provided by node $B$.
In this way, nodes in $Gamma$ offer lower latency than nodes in $A_{t(1)}$.

Services from the concurrent class are migrated to nodes of different tiers if, and only if, the storage capacity and latency offered are appropriate for such services.
Otherwise, nodes in $Y$ are not reserved.

\section{Improving real-time traffic forecast}
\label{sec:traffic-forecast}
Most time series prediction methods are based on the assumption that past observations contain all the information about the pattern of behavior of the time series, and that pattern is recurrent over time.
In the literature, numerous methods describe the behavior of time series, and they are divided into two approaches: classical and learning~\cite{liu2017survey}.
On the one hand, the classical approach about the Exponential Smoothing and \gls{arima} methods, which use parametric statistics to transform their dataset into a known probability distribution, and therefore need to know the behavior of the time series~\cite{granger2014forecasting}.
On the other hand, deep learning models such as \gls{gru} and \gls{rnn} do not depend on prior knowledge of time series properties~\cite{andrews1995survey}.
These models are simpler to be adjusted and demonstrate considerable performance even when applied to complex and highly non-linear series.

In this work, two models are designed, based on a classic parametric modeling \gls{arima} and a deep neural network architecture \gls{lstm}, namely \gls{arima-pred} and \gls{lstm-pred}, respectively, to solve the traffic forecasting problem.
The comparisons with other baselines show the effectiveness of these methods~\cite{nagy2018survey, 8667446}.
Besides, they are widely used for real-time predictions, such as forecasts of network traffic flows, been used in several studies~\cite{nagy2018survey, zhang2018citywide, zhang2019deep}. 
In practice, the development of each regressive model differs (i) in the modeling of the time series for a set appropriate to the model (ii) and the choice of parameters most appropriate for each type of approach~\cite{prettenhofer2014gradient}.
Performance evaluation for both models are discussed in Section~\ref{sec:evaluations-and-experiments}.

\subsection{ARIMA-PRED}
\label{sec:ARIMA-PRED}

ARIMA model is used to understand time series or predict a point in the future.
Any time series that exhibits patterns and is not random white noise can be modeled with ARIMA models~\cite{zhang2003time}.

The Autoregressive part (\textbf{AR}) of the method indicates that the variable of interest undergoes a regression on its previous values.
The integrated part (\textbf{I}) indicates that the data values have been replaced with the difference between their current and previous values, making the series stationary (this process can be performed more than once).
The Moving Average part (\textbf{MA}) indicates that the regression error is a linear combination of the error terms applied to past observations.
The purpose of each component is to make the model fit the data as best as possible~\cite{contreras2003arima}.

Non-seasonal ARIMA models are usually denoted by ARIMA\textbf{(p, q, d)}, where:
\begin{itemize}
\item \textbf{p}: number of lags of the autoregressive model.
\item \textbf{d}: number of times the data has had past values subtracted.
\item \textbf{q}: order of the moving average model.
\end{itemize}

The model can be written as follows:
\begin{align}
        y_{t} = \alpha + \phi_{1} y_{t-1} + \phi_{2} y_{t-2} + ... + \phi_{p} y_{t-p} +\notag\\\varepsilon{t} -\theta_{1} \varepsilon_{t-1} - \theta_{2} \varepsilon_{t-2} - ... - \theta_{q} \varepsilon_{t-q}    
\end{align}

where $\alpha$ is a constant, $\phi$ is an estimated coefficient, and $y$ deals with past entries (\textit{lags});
$\theta $ is an estimated coefficient and $\varepsilon$ is the errors associated with past regressive predictions.

ARIMA models generally perform best when the series is relatively long and well behaved.
If the series is very irregular, the results are generally inferior to those obtained by other methods, such as recurrent neural networks.

\subsection{LSTM-PRED}
\label{sec:LSTM-PRED}

The \gls{lstm-pred} is based on \gls{lstm} model, which processes data passing on information as it propagates forward.
This model can recognize long-term patterns and dependencies, ideal for classifying, processing and predicting time series with time intervals of unknown duration.
This is possible due to the ability to remove or add information to the state of the cell, regulated by structures called gates~\cite{hunt1992neural}.

The cell's state, in theory, acts as a pathway that carries relevant information along the entire sequence chain.
Information is added to or removed from the cell state through gates, which decide what information is allowed in the cell state.
They learn what information is relevant to keep or forget during training.
In general, models based on \gls{lstm} networks has three gates:

\begin{itemize}
\item \textbf{Forget Gate}: removes information that is not useful to the cell state.
\item \textbf{Input Gate}: adds useful information to cell state.
\item \textbf{Output Gate}: extracts useful information from the current cell state to decide what the next hidden state should be.
\end{itemize}

\begin{figure}[]
    \centering
        \includegraphics[width=.6\linewidth]{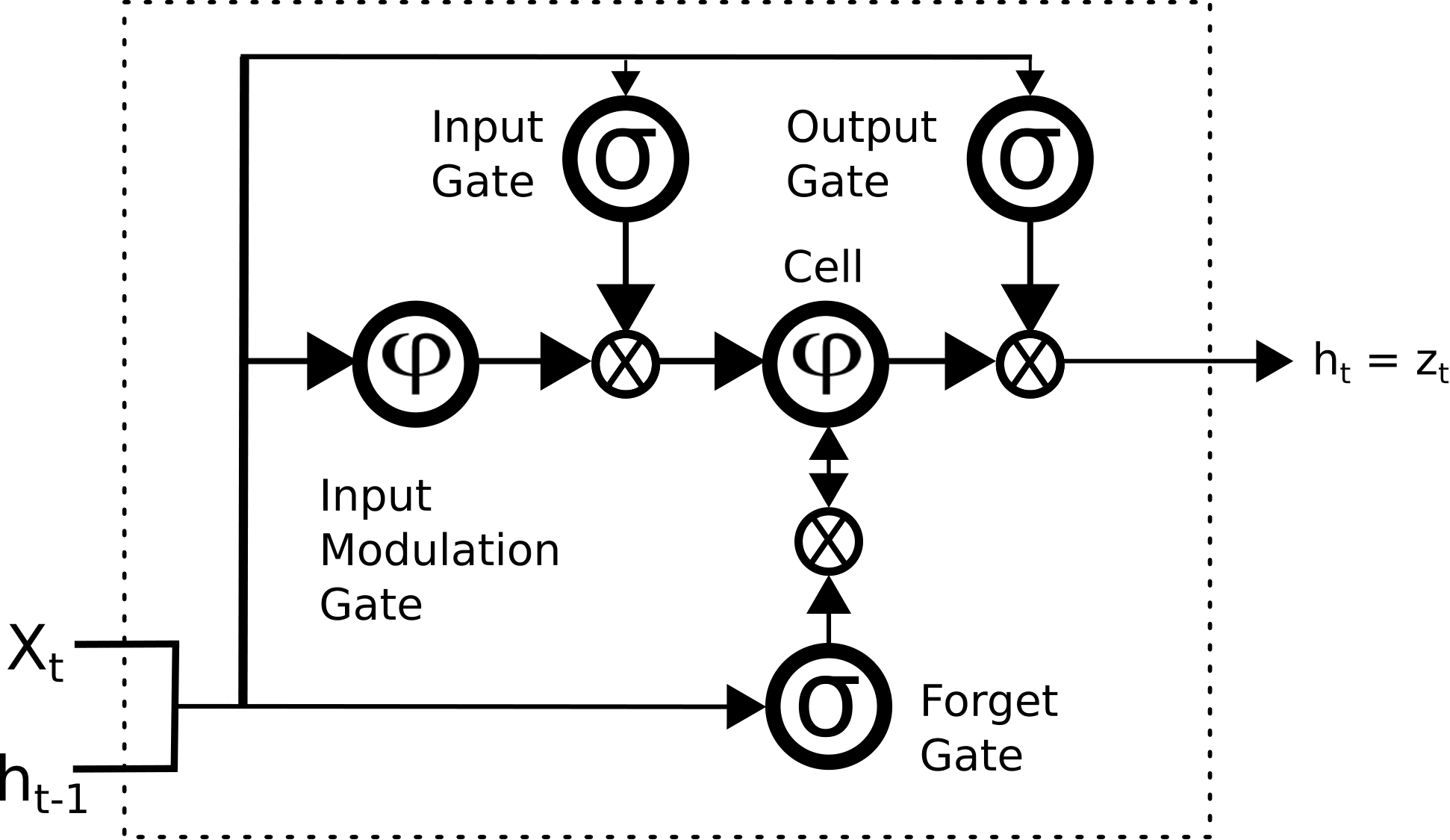}
    \caption{\gls{lstm} cell.}
\label{fig:LSTM-cell}
\end{figure}

Figure\ref{fig:LSTM-cell} shows the structure of the \gls{lstm} cell.

\gls{lstm} networks can be applied to a variety of deep learning tasks, which primarily include prediction based on prior information.
Examples include text prediction, business actions, and network traffic volume.

\section{Evaluations and experiments}
\label{sec:evaluations-and-experiments}
We show the performance of the \gls{tiptop} using a real-world mobile network traffic data in Milan, Italy\cite{barlacchi2015multi}. 

Section~\ref{subsec:scenarios} describes the scenario used in this work. Section~\ref{sec:method-application} shows the design and implementation of the method proposed to build an environment based on Cloud-Fog Computing.
Section~\ref{sec:traffic-intensity} discusses the development of the \gls{arima-pred} and \gls{lstm-pred} models.
Section~\ref{sec:perfomance-prediction-models} shows the performance assessment of the \gls{smart-fl} considering six snapshots selected in particular days and hours based on the traffic intensity.
Finally, Section~\ref{sec:traffic-predicted} shows the performance assessment of the \gls{smart-fl} and \gls{tiptop} compared with two algorithms considering one month of the predicted network traffic.

\subsection{Scenarios description}
\label{subsec:scenarios}
\begin{figure}[!htb]
    \centering
    \begin{subfigure}[]{.3\textwidth}
        \includegraphics[width=1\textwidth]{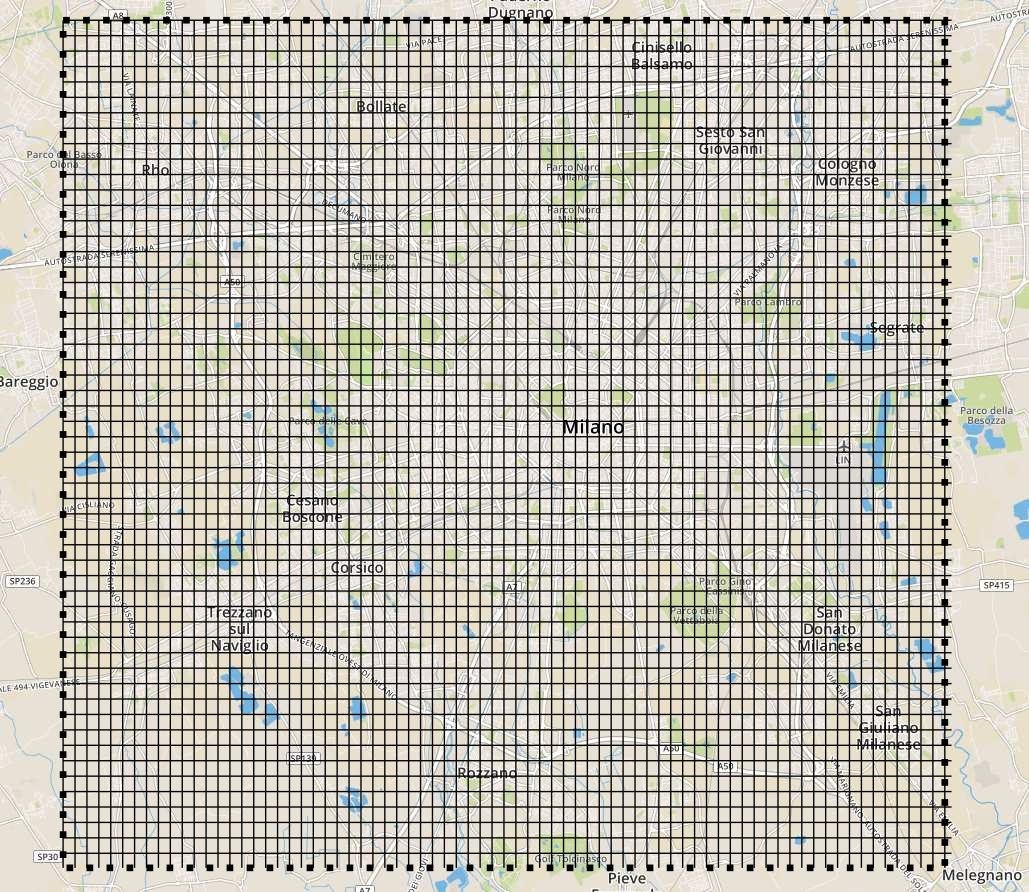}
        \caption{Milan grid.}
        \label{fig:scenario-grid}
    \end{subfigure} 
    \begin{subfigure}[]{.3\textwidth}
        \includegraphics[width=1\textwidth]{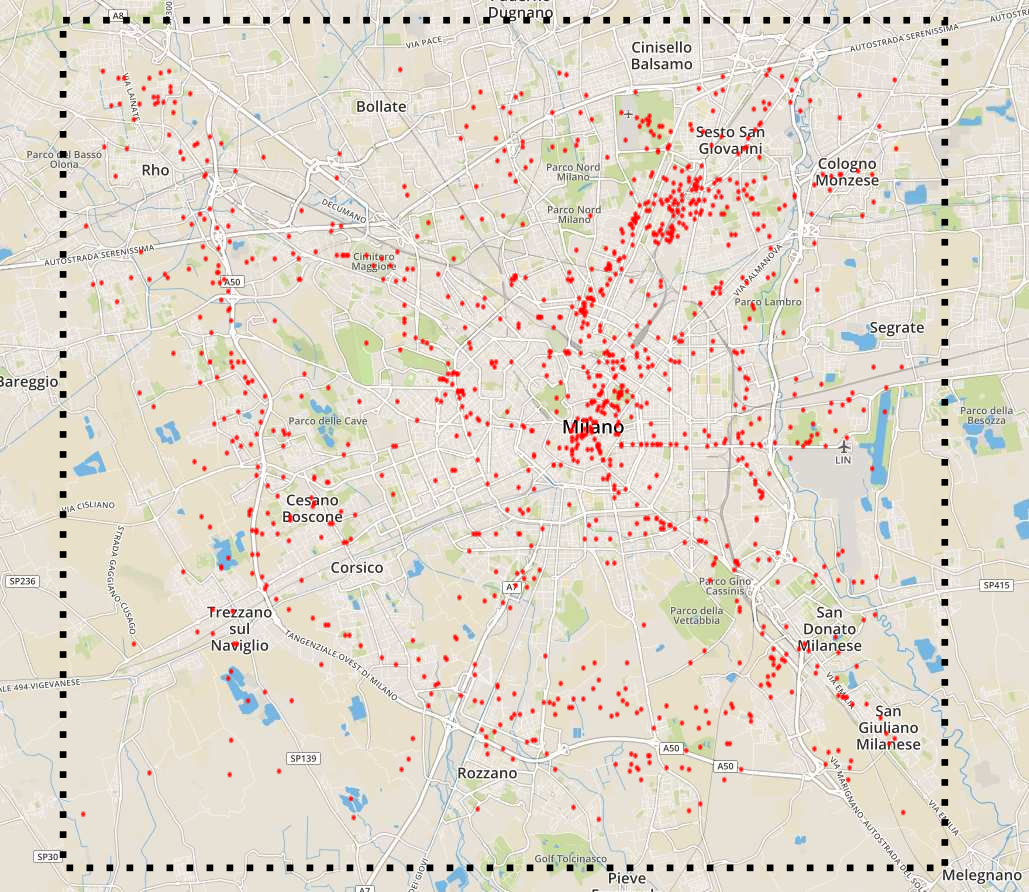}
        \caption{Base station deploys in Milan.}
        \label{fig:scenario-base-station}
    \end{subfigure} 
     \caption{Map view of the scenarios studied.}
     \label{fig:scenario_raw}
\end{figure}

We experiment with publicly available real-world mobile traffic data sets, which contain two months of network traffic data~(November/2013 to December/2013) released through Telecom Italia’s Big Data Challenge~\cite{barlacchi2015multi}.
The unique multi-source composition of the dataset makes it an ideal dataset to analyze various problems, including energy consumption, mobility planning, event detection, and many others.
Also, it is the richest open multi-source data set ever released on two geographical areas.
Figure~\ref{fig:scenario_raw} illustrates this scenario.

The geographical area is composed of a 100 $\times$ 100 grid, with a size of 235 $\times$ 235 $m^2$ each, illustrated in Figure~\ref{fig:scenario-grid}.
Every time a mobile user requests services to a telecommunication provider, a \gls{cdr} is recorded.
This information is then compiled into 10-minute intervals.
Furthermore, a base station set $BS = \{bs_{1},\;bs_{2},\;...,\;bs_{bs}\}$ was obtained from CellMapper\footnote{https://www.cellmapper.net/map}, which consists of the locations and coverage areas of active base stations observed in the two months periods, illustrated in Figure~\ref{fig:scenario-base-station}.
The grids were mapped to the base stations' coverage areas and aggregated the \gls{cdr} amount per base station. 
It is considered that there are multimedia services requests in a grid if its \gls{cdr} amount is above the average. 
The multimedia services requests are aggregated by region.

Table~\ref{tab:dataset-lines} shows the first five values of network traffic in cell $\#1$.
The first column (\textit{timestamp}) represents a timestamp variable, starting on 2013-11-01 00:10:00 and ending on 2014-01-01 23:40:00.
The second column (\textit{traffic volume}) represents the amount of \gls{cdr} generated.

Given the temporal organization, these data can be modeled and evaluated as a time series, where the neighboring observations are dependent, and the interest is to analyze and model this dependency.

\begingroup
\renewcommand\arraystretch{1}
\begin{table}[H]
\caption[The dataset structure]{The dataset structure.}
\centering
\scalebox{0.9}{
\begin{tabular}{@{}ccc@{}}
\toprule
 & \textbf{timestamp} & \textbf{traffic volume}\\ 
\midrule 
\textbf{0} & 2013-11-01 00:00:00 & 11.028366381681\\ 
\textbf{1} & 2013-11-01 00:10:00 & 11.1271008756737\\ 
\textbf{2} & 2013-11-01 00:20:00 & 10.8927706027911\\ 
\textbf{3} & 2013-11-01 00:30:00 & 8.62242459098975\\ 
\textbf{4} & 2013-11-01 00:40:00 & 8.00992746244576\\ 
\textbf{...} & ... & ...\\ 
\bottomrule
\end{tabular}}
\label{tab:dataset-lines}
\end{table}
\endgroup

\begin{figure}[!htb]
    \centering
    \begin{subfigure}[]{.3\textwidth}
        \includegraphics[width=1\textwidth]{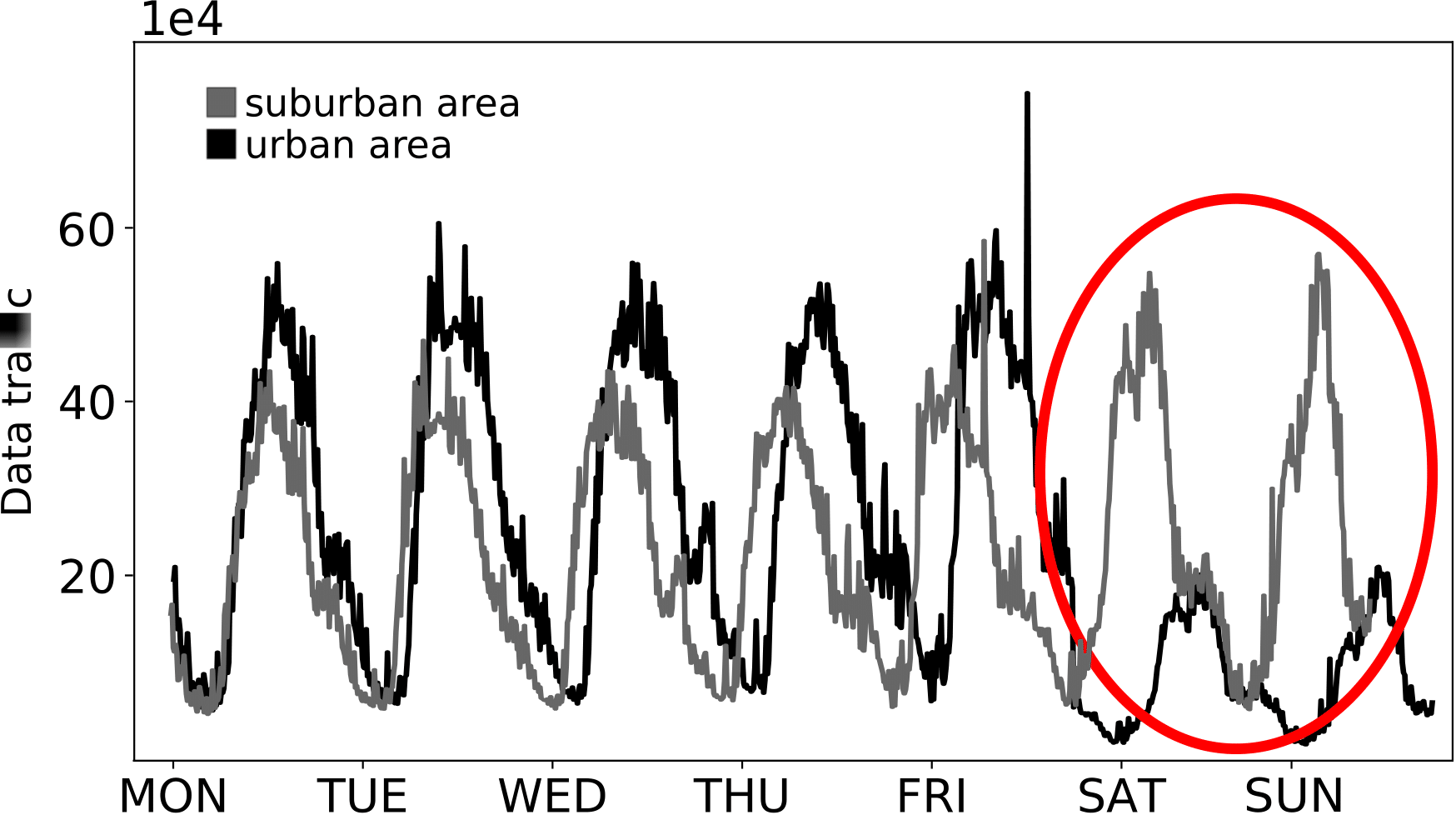}
        \caption{Average traffic volume over one week (urban and suburban area).}
        \label{fig:traffic-one-week}
    \end{subfigure} 
    \begin{subfigure}[]{.3\textwidth}
        \includegraphics[width=1\textwidth]{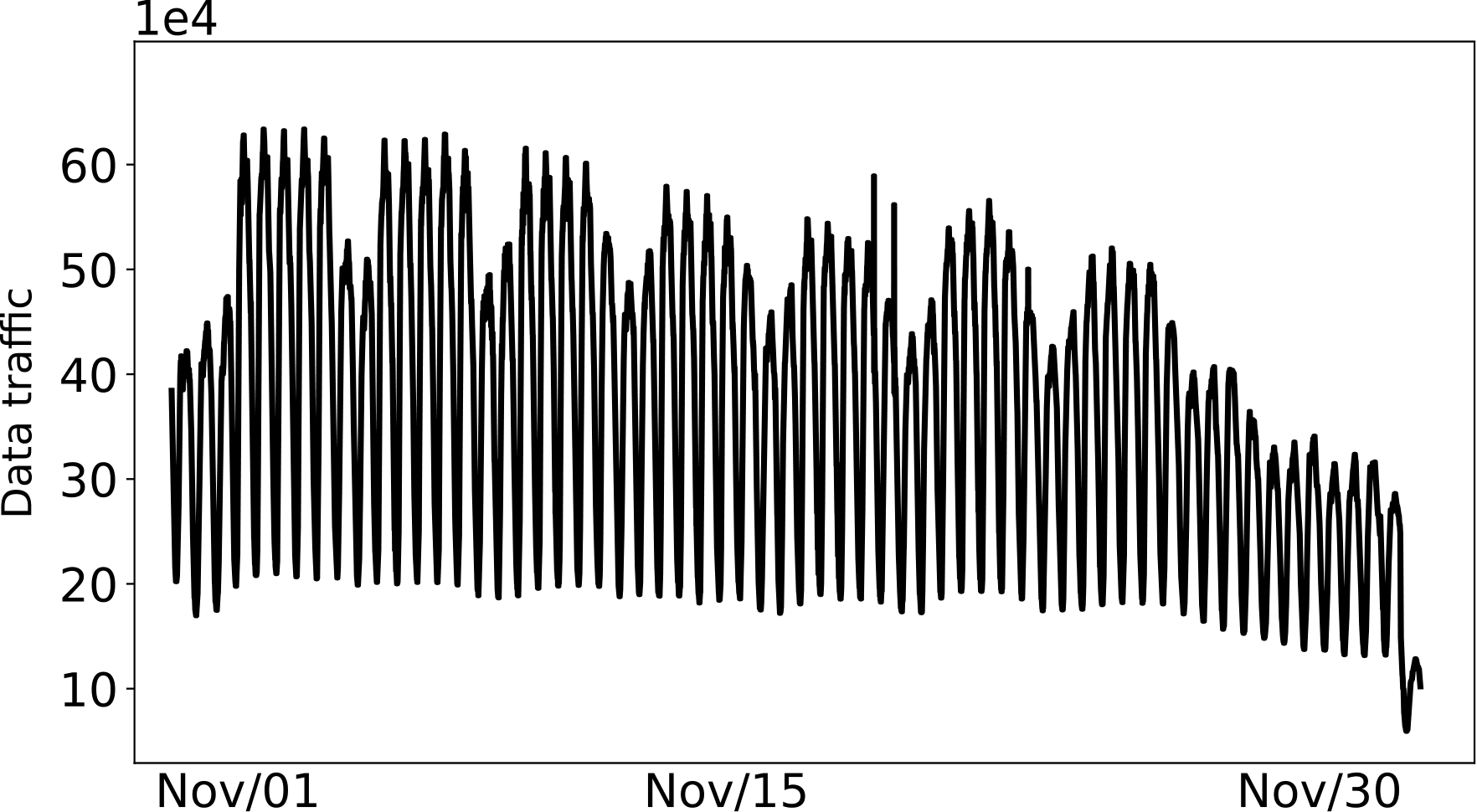}
        \caption{Average traffic volume over one month.}
        \label{fig:traffic-one-month}
    \end{subfigure} 
     \caption{Average traffic volume of all base stations.}
     \label{fig:traffic-volume}
\end{figure}

Figure~\ref{fig:traffic-volume} shows the trends in three different traffic flows over a seven and thirty days period, both with 10 minutes granularity.
Based on Figure~\ref{fig:traffic-one-week}, the traffic data is higher during the working hours than at midnight and lower on weekends than on weekdays.
The picture clearly shows that the data traffic exhibits certain periodicity (in daily and weekly patterns) due to regular working schedules.
On the one hand, the urban area has lower traffic data on weekends than during the week due to urban mobility and regular working hours.
On the other hand, the suburban area has regular traffic data during the seven days of the week.
Figure~\ref{fig:traffic-one-month} shows the average traffic volume over one month of November.

\subsection{Proposed method application}
\label{sec:method-application}

The design and implementation of the method proposed to build an environment based on Cloud-Fog Computing were applied to the existing cellular network described earlier.
However, it can also be applied to other datasets.

The method starts with the definition of the set $BS$ and the parameters $\Xi$ and $\mu$.
$BS$ represents the base stations in the scenario described above;
$\Xi$ is defined as the radius $r\;=\;3km$;
and $\mu\;=\;2$, but can be any value.
Since $\Xi$ is defined as the radius $r\;=\;3km$, this means that there is an edge between $bs_{i}$ and $bs_ {j}$ $\in\;BS$, if and only if, the distance between $bs_{i}$ and $bs_{j}$ is $3km$.
It is worth mentioning that, in this step, other forms of connection are possible, such as signal strength or traffic similarity.

\begin{figure*}[!htb]
    \centering
    \begin{subfigure}[]{.32\textwidth}
        \includegraphics[width=1\textwidth]{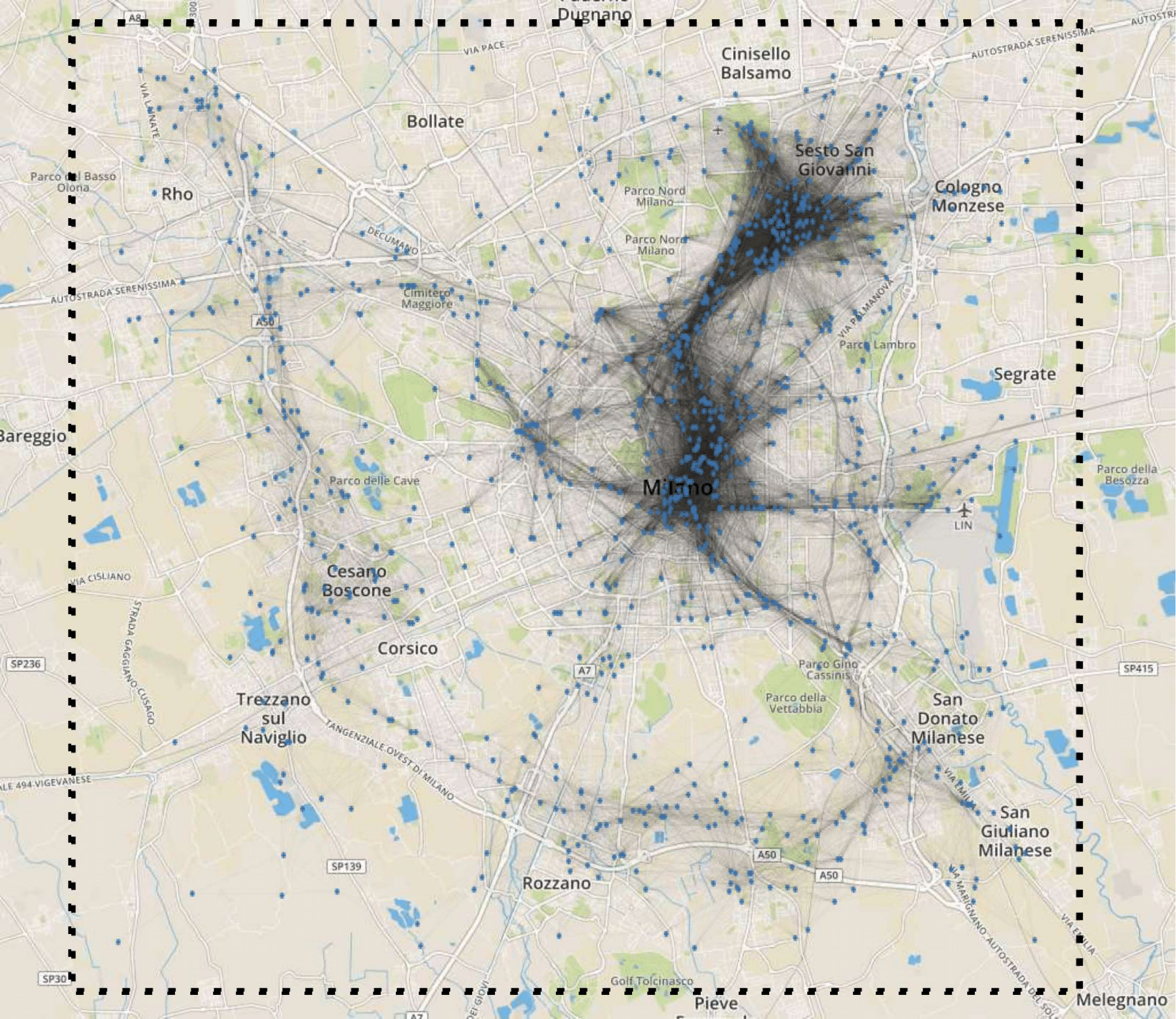}
        \caption{Graph $\mathcal{G}$ (base stations in blue).}
        \label{fig:graph-bs}
    \end{subfigure} 
    \begin{subfigure}[]{.32\textwidth}
        \includegraphics[width=1\textwidth]{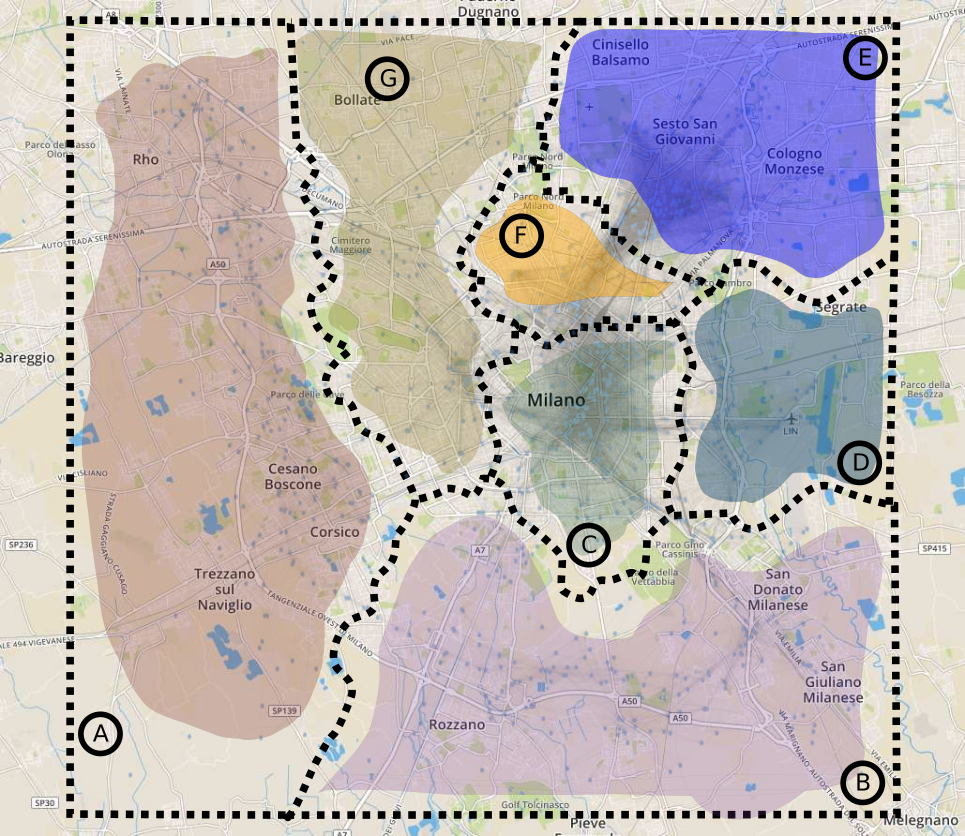}
        \caption{Communities detected by Louvain.}
         \label{fig:graph-comunidades}
    \end{subfigure} 
    \begin{subfigure}[]{.32\textwidth}
        \includegraphics[width=1\textwidth]{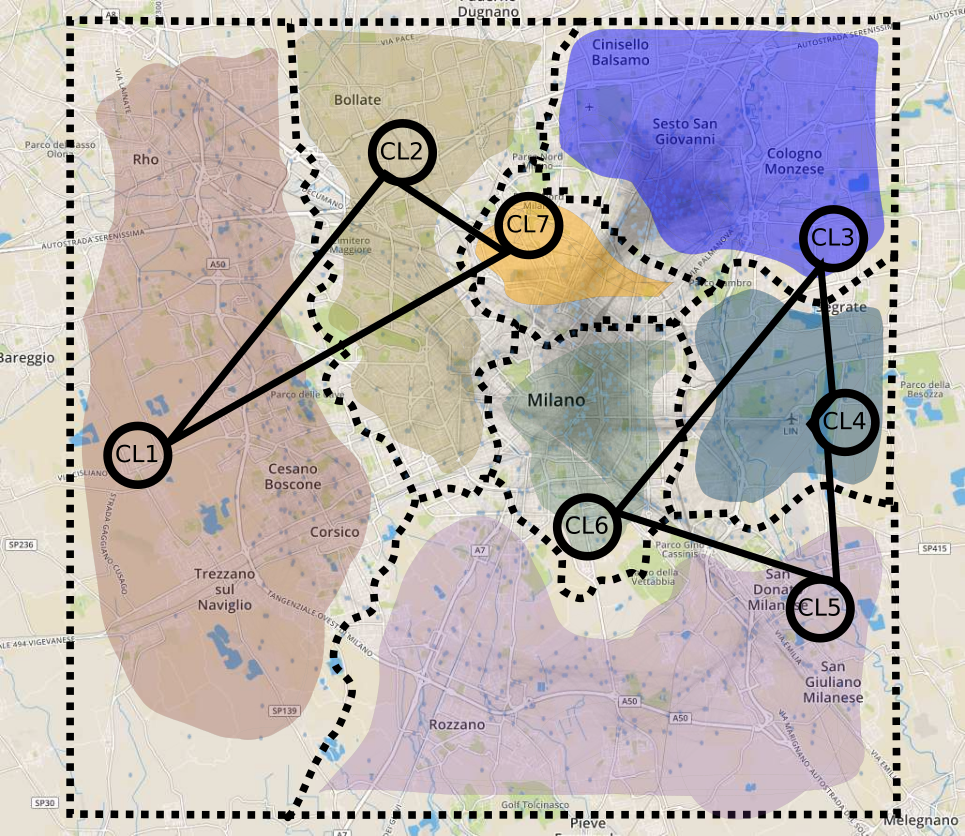}
        \caption{Seven cloudlets nodes.}
        \label{fig:graph-cloudlets}
    \end{subfigure} 
    \begin{subfigure}[]{.32\textwidth}
        \includegraphics[width=1\textwidth]{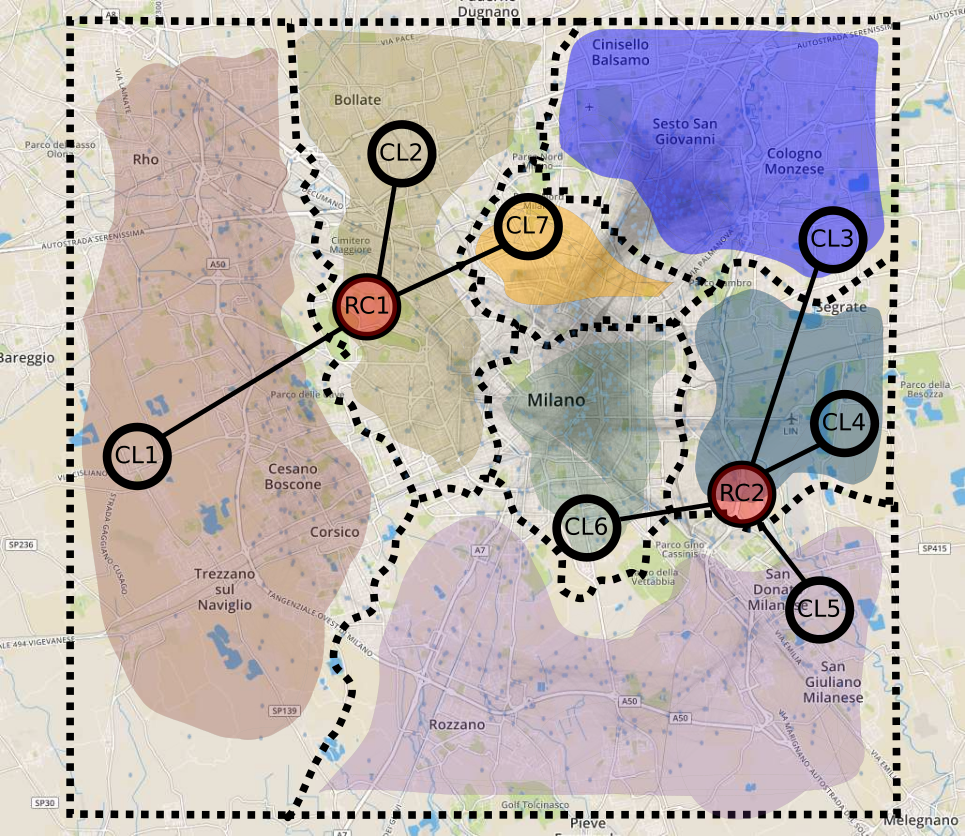}
        \caption{Two regional cloudlets nodes.}
        \label{fig:graph-regional-cloud}
    \end{subfigure} 
    \begin{subfigure}[]{.32\textwidth}
        \includegraphics[width=1\textwidth]{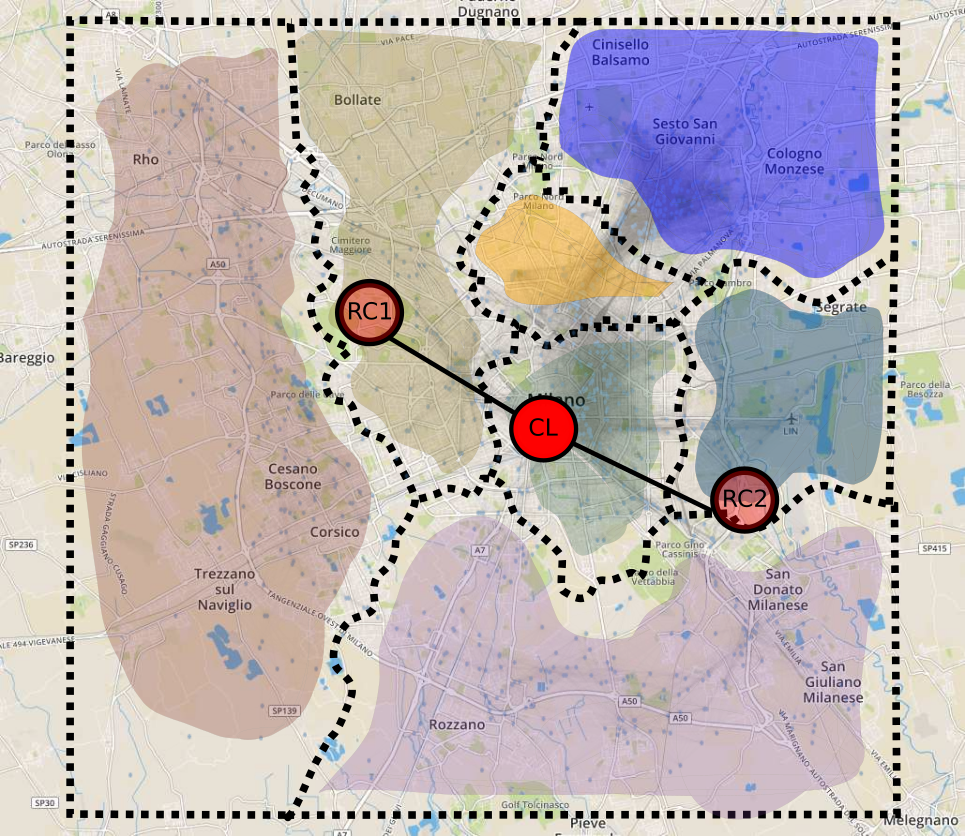}
        \caption{One Cloud node.}
        \label{fig:graph-cloud}
    \end{subfigure} 
    \begin{subfigure}[]{.32\textwidth}
        \includegraphics[width=1\textwidth]{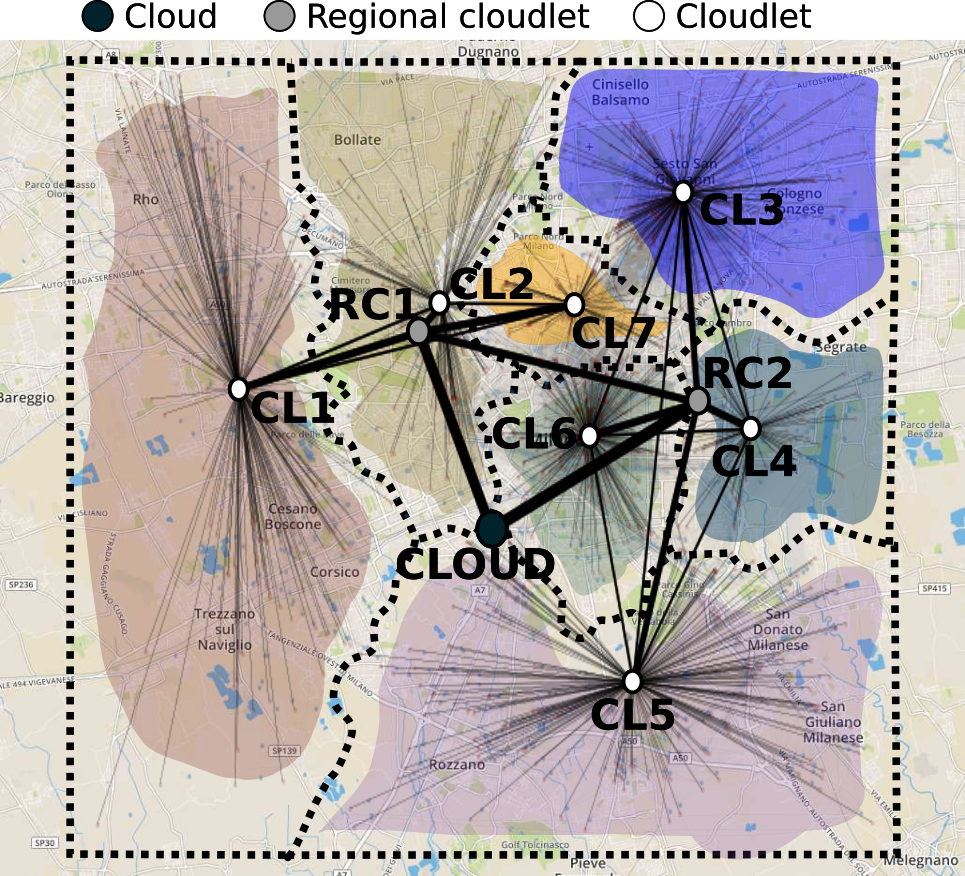}
        \caption{Map view of the network topology.}
         \label{fig:graph-final}
    \end{subfigure} 
     \caption{Designing of Cloud-Fog Hierarchical Environments.}
     \label{fig:scenarios-all}
\end{figure*}

Figure~\ref{fig:scenarios-all} illustrates the development of the hierarchical scenario resulting from each step performed.
Figure~\ref{fig:graph-bs} illustrates the graph $\mathcal {G} = (BS, \mathcal {E})$ modeled from step~\textbf{1}.
Figure~\ref{fig:graph-comunidades} depicts the seven communities over the base stations (colored dots), represented by the letters A to G, from step~\textbf{2}.
Each community portrays a region.
It is possible to notice that many regions (e.g., C, D, F, and G) are composed of an urban and suburban segment.
This indicates that the base stations in these areas are potentially complementary due to traffic patterns.
The Louvain heuristic is used to find the community set $S$.
It is a fast algorithm $O(n+m \cdot logn+m)$, where $n$ and $m$ are the numbers of vertices and edges, respectively, to detect communities in large-scale networks based on modularity optimization~\cite{blondel2008fast}.
This method aims to find partitions (structures composed of communities) that maximize the density of intra-group connections concerning the density of inter-group connections and find dense optimal sub-graphs in large graphs.

The environment obtained from the steps (\textbf{3-5}), the addition of an upper-tier node for each community detected, the addition of edges between all nodes in the current tier, and the detection of communities and removal of edges between nodes of different communities, respectively, are illustrated in Figure~\ref{fig:graph-cloudlets}.
The step (\textbf{6}), adding an upper-tier node with links between the node and the subgraph for each subgraph found, is illustrated in Figure ~\ref{fig:graph-regional-cloud}.
The stopping criterion is $\mu\;=\;2$ and must be observed at each step.
In this case, the stopping criterion is met, adding an upper-tier node that connects to the lower tier nodes, as shown in Figure~\ref{fig:graph-cloud}.
Figure~\ref{fig:graph-final} illustrates the final Cloud-Fog hierarchical environment.

It is considered that nodes added of each tier from steps \textbf {(2)} to \textbf {(6)} are labeled cloudlet (\textbf {CL1}, \textbf{CL2}, \textbf{CL3}, \textbf{CL4}, \textbf{CL5}, \textbf{CL6} and \textbf{CL7}), regional Cloud (\textbf{RC1} and \textbf{RC2}) and Cloud (\textbf{CL}), respectively.
Table~\ref{table:scenario-description} shows the numbers of nodes and average coverage areas per tier.

\begingroup
\renewcommand\arraystretch{1.5}
\begin{table}[!htb]
\caption{Number of nodes and average coverage area per tier.}
\centering
\scalebox{0.7}{
\begin{tabular}{@{}ccc@{}}
\toprule
\textbf{Nodes} & \textbf{Number of nodes} &  \textbf{Coverage areas ($m^{2}$ per node)}\\ 
\midrule 
Base station & 1150  & 492.46 \\
Cloudlets & 7  & 9072.67 \\ 
Regional Cloud & 2 & 31754.37 \\ 
Cloud & 1 & 552250 \\
\bottomrule
\end{tabular}}
\label{table:scenario-description}
\end{table}
\endgroup

Service requests belong to one of two classes in this environment: multimedia or concurrent.
Let $W$ be the set that represents the services from the multimedia class and $J$ be the set that represents the services from the concurrent class.
On the one hand, multimedia class' services are \gls{vod}, interactive video 3D, high-definition, \gls{uhd} (that includes 4K UHD and 8K UHD) video streaming, etc.
On the other hand, concurrent class' services are offered in \gls{tcs} and \gls{iot}.
Thus, the network and nodes' resources, such as bandwidth and hardware capacity, are competed by both classes' services.
The main idea is to create the most competitive environment.

Let $Lat\;=\;\{lat_{1},\;lat_{2},\;lat_{3},\;...,\;lat_{r}\}$ and $Mip\;=\;\{mip_{1},\;mip_{2},\; mip_{3},\;...,\;mip_{r}\}$ be the sets that represents the maximum latency and \gls{mips} required for each service $w\;\in\;W$ and $j\;\in\;J$, based on~\cite{gupta2017ifogsim, sinqadu2020performance, perala2018fog}.
Also, $V_t\;=\;\{v_ {1},\;v_ {2},\;v_ {3},\;...,\;v_ {g}\} $ be the set that represents the network traffic in time $t$ for each grid $g$.
Thus, the service requirements are modeled for both classes in terms of three parameters:
maximum service latency ($lat_{i}$ - in milliseconds);
maximum amount of RAM for temporary storage ($mip_{i}$ - in $GB$);
and maximum amount of \gls{mips} ($v_g^t$).
Therefore, for each $w\;\in\;W$ and $j\;\in\;J$:

\begin{align}
    \MoveEqLeft \displaystyle w_{i} = \{lat_{i},\;mip_{i},\;v_g^t\},\\
    \MoveEqLeft \displaystyle j_{i} = \{lat_{i},\;mip_{i},\;v_g^t\}
\end{align}

where, $0\;\leq\;i\;\leq\;r$, $lat_{i}\;\in\;Lat$, $mip_{i}\;\in\;Mip$, and $v_g^t\;\in\;V_t$.
The maximum storage required to perform any service requested at time $t$ in the grid $g$ is $v_g^t$, where $v_g^t\;\in\;V_t$.
In that way, the storage required to perform any services varies according to network traffic, showing periodicity.

The nodes capable of storing and processing services are modeled in terms of five parameters:
\gls{mips} available ($mips$);
storage available ($stor$ - in $GB$);
RAM available ($ram$ - in $GB$);
$uplink$ ($up\_rate$) and $downlink$ ($down\_rate$) rates offered, both in $Mb$.
Table~\ref{table:parametros-nos} shows the range of values for each parameter, based on~\cite{gupta2017ifogsim}.
Therefore, for each $ell\;\in\;\La$:

\begin{equation}
\ell = \{mips,\;stor,\;ram,\;up\_rate,\;down\_rate\}
\end{equation}

\begingroup
\renewcommand\arraystretch{1.5}
\begin{table}[!htb]
\caption{Range of values to model the nodes by tiers.}
\centering
\scalebox{0.8}{
\begin{tabular}{@{}ccccc@{}}
\toprule
\textbf{Parameters} & \multicolumn{4}{c}{\textbf{Values per tier}}\\ 
 & $1^{\circ}$ & $2^{\circ}$ & $3^{\circ}$ & $4^{\circ}$\\
\midrule 
$mips$ & [2.8 - 5.3] & [5.3-7.8] & [7.8-10.2] & [10.2-20.5]\\
$stor$ & $100^2$ & $200^2$ & $400^2$ & $1000^2$\\
$ram$ & 25 & 40 & 60 & 100\\ 
$up\_rate$ \& $down\_rate$ & 300 & 500 & 800 & 2000\\
\bottomrule
\end{tabular}}
\label{table:parametros-nos}
\end{table}
\endgroup

It is worth noticing that all parameters, such as $lat_i$, $mip_i$, $mips$, $stor$, $ram$, $up\_rate$, and $down\_rate$, change dynamically over time~\cite{puliafito2020mobfogsim}.
Besides, higher-tier nodes have more network and hardware resources, such as bandwidth and hardware capacity, than lower-tier nodes.
In contrast, lower-tier nodes support latency-sensitive services, not fully supported by \gls{cc}.
The performance assessment was conducted in a simulated environment on MultiTierFogSim\footnote{https://github.com/fillipesansilva/MultiTierFogSim.git}.

\subsection{Network traffic intensity}
\label{sec:traffic-intensity}
\begin{figure}[H]
    \centering
    \includegraphics[width=0.7\linewidth]{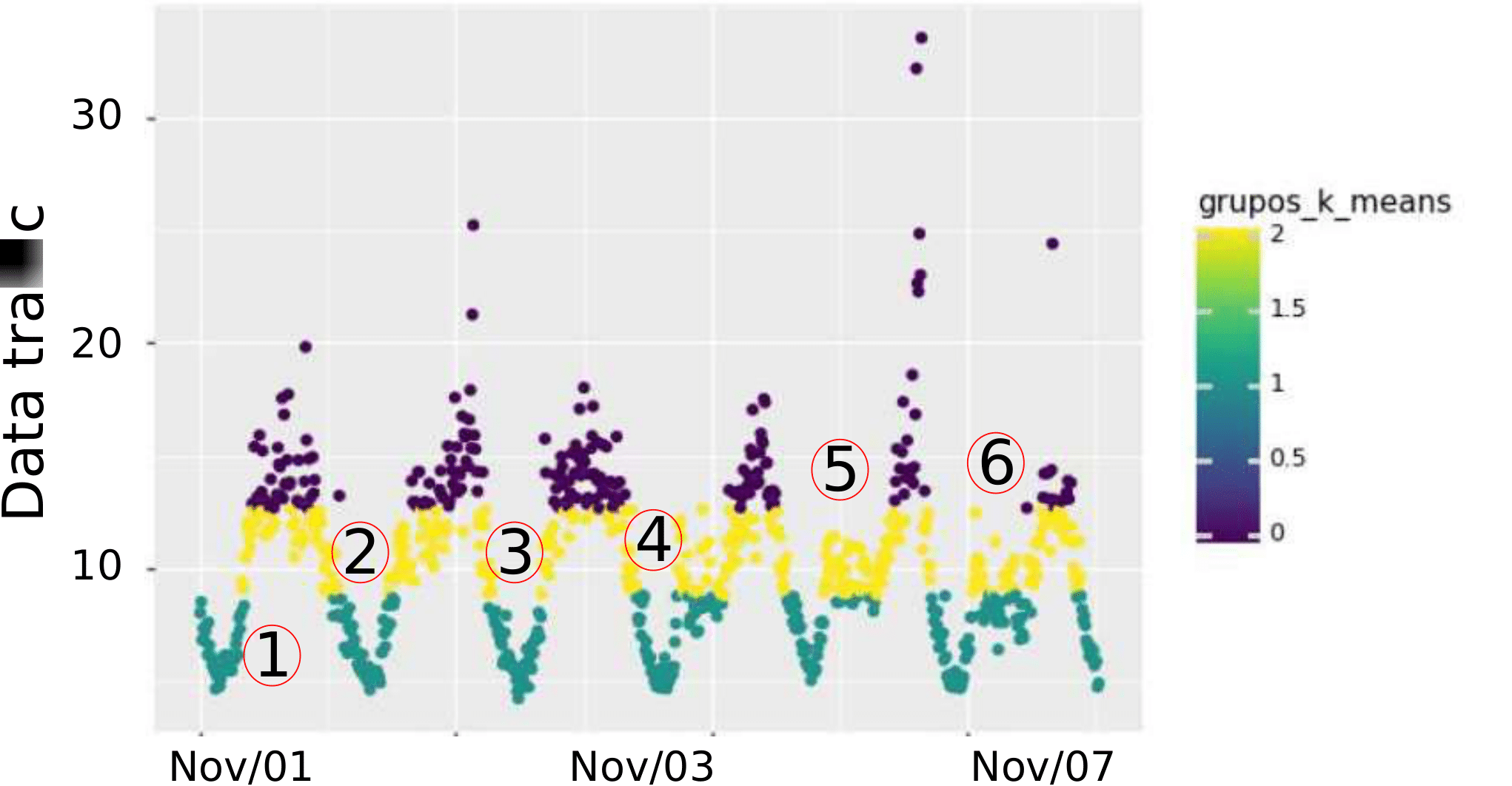}
    \caption{K-Means clustering in network traffic.}
\label{fig:kmeans}
\end{figure}

Six snapshots selected in particular days and hours based on the network traffic intensity are analyzed to assess the performance of the \gls{smart-fl} algorithm in different circumstances in Milan - Italy. 
The main idea is to evaluate the positioning of the selected nodes in relation to their storage capacity, latency offered, and the demands' geographic location for multimedia services.
To do that, it is applied the \textbf{K}-means algorithm to partition the network traffic intensity into 3-predefined distinct non-overlapping subgroups where each data point belongs to only one group.
The network traffic intensity is grouped into low, medium, and high traffic intensity.
Figure~\ref{fig:kmeans} illustrates a part of the training set during the first week of November.

\begin{figure*}
  \centering
  \begin{minipage}{.3\linewidth}
    \centering
    \subcaptionbox{Multimedia requests and selected nodes.\label{fig:request1}}
      {\includegraphics[width=\linewidth]{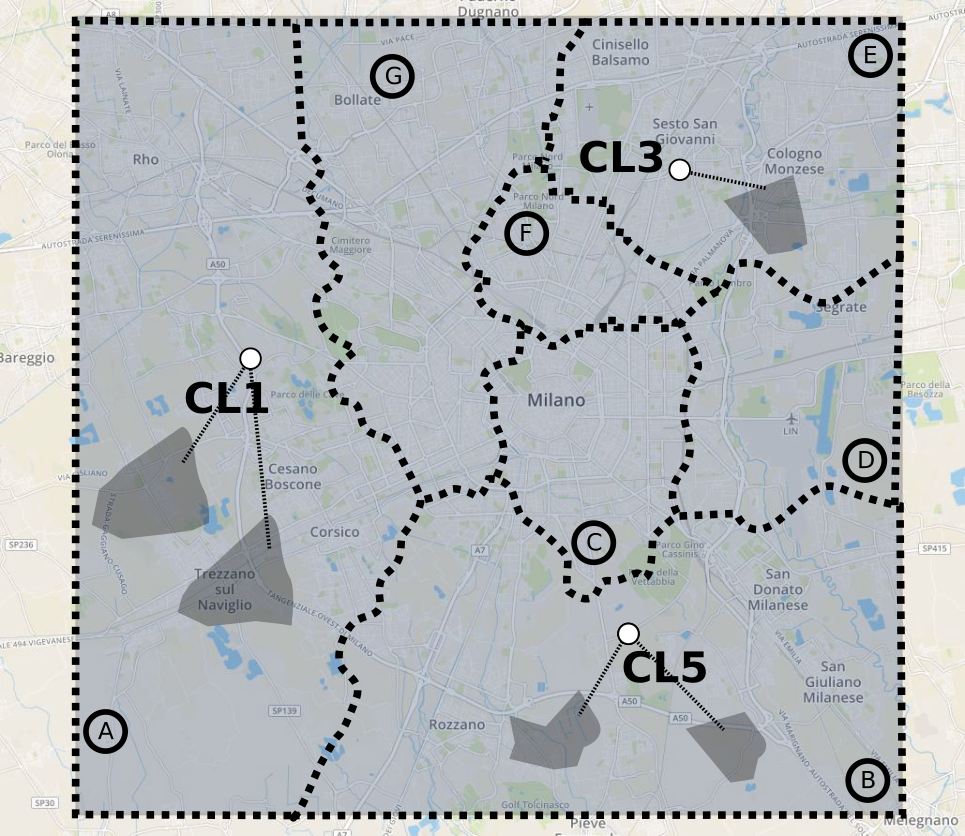}}
    \subcaptionbox{Nodes' hardware capacity and latency.\label{fig:fog_nodes_capacity_latency1}}
      {\includegraphics[width=\linewidth]{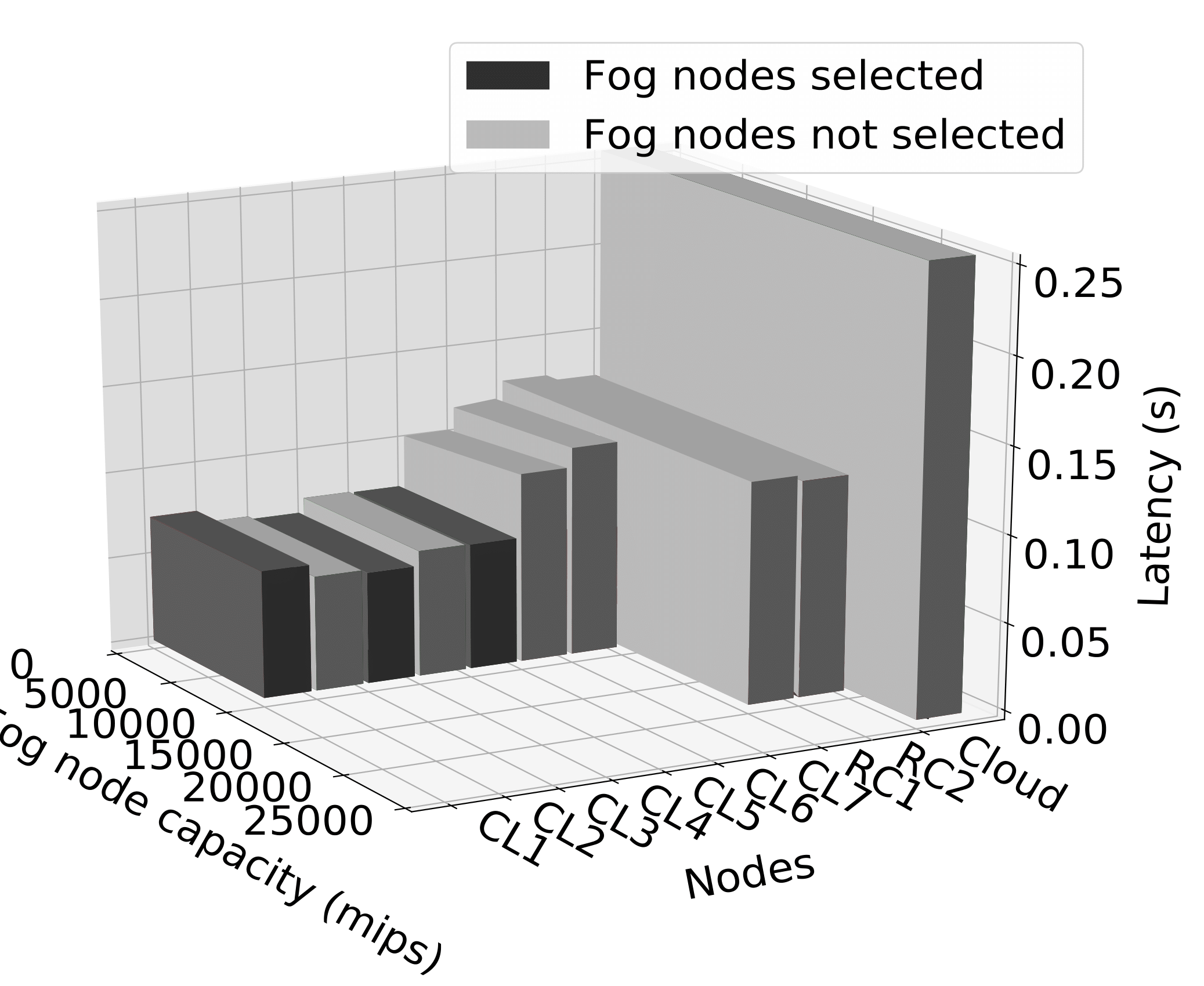}}
    \caption{Low traffic intensity.}
    \label{fig:region1}
  \end{minipage}
  \quad
  \begin{minipage}{.3\linewidth}
    \centering
    \subcaptionbox{Multimedia requests and selected nodes.\label{fig:request2}}
      {\includegraphics[width=\linewidth]{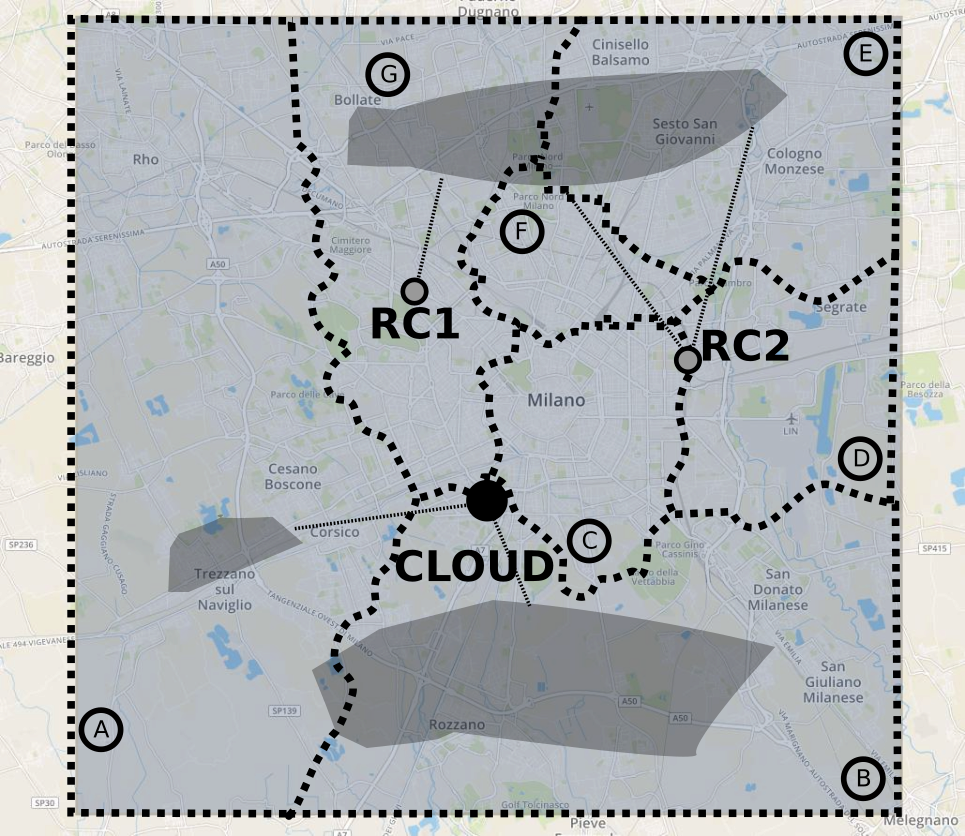}}
    \subcaptionbox{Nodes' hardware capacity and latency.\label{fig:fog_nodes_capacity_latency2}}
      {\includegraphics[width=\linewidth]{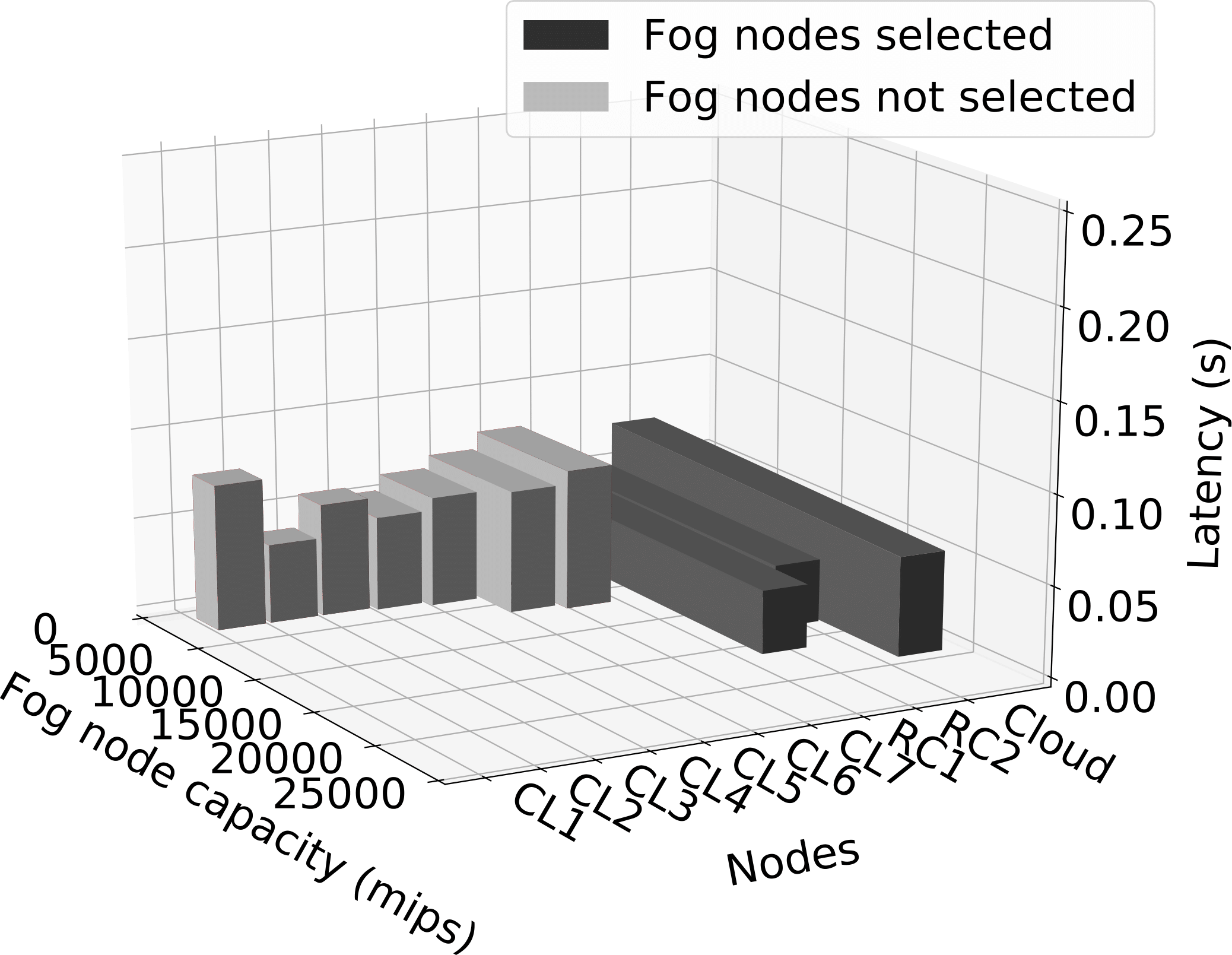}}
    \caption{Medium traffic intensity.}
    \label{fig:region2}
  \end{minipage}
  \quad
  \begin{minipage}{.3\linewidth}
    \centering
    \subcaptionbox{Multimedia requests and selected nodes.\label{fig:request3}}
      {\includegraphics[width=\linewidth]{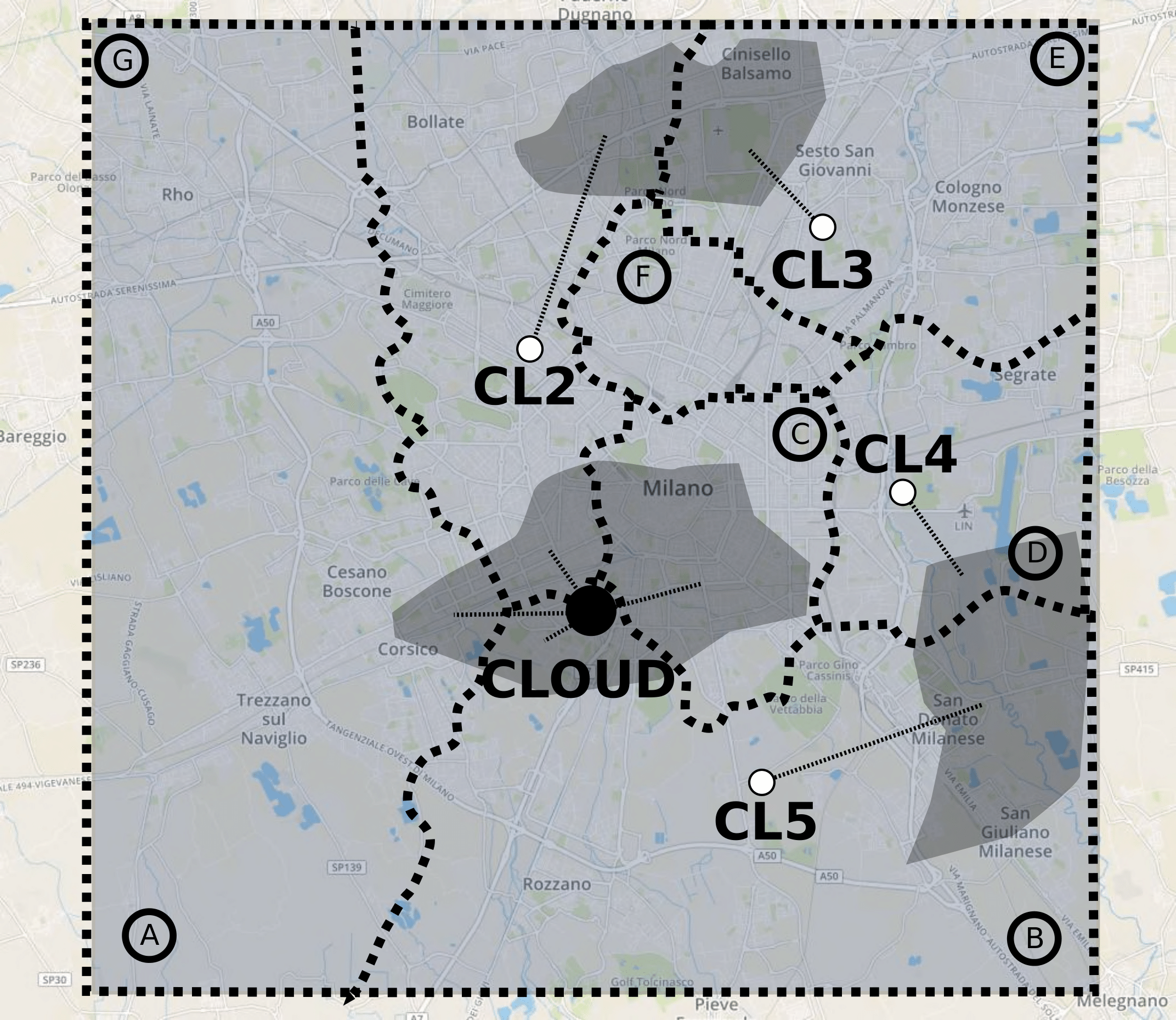}}
    \subcaptionbox{Nodes' hardware capacity and latency.\label{fig:fog_nodes_capacity_latency3}}
      {\includegraphics[width=\linewidth]{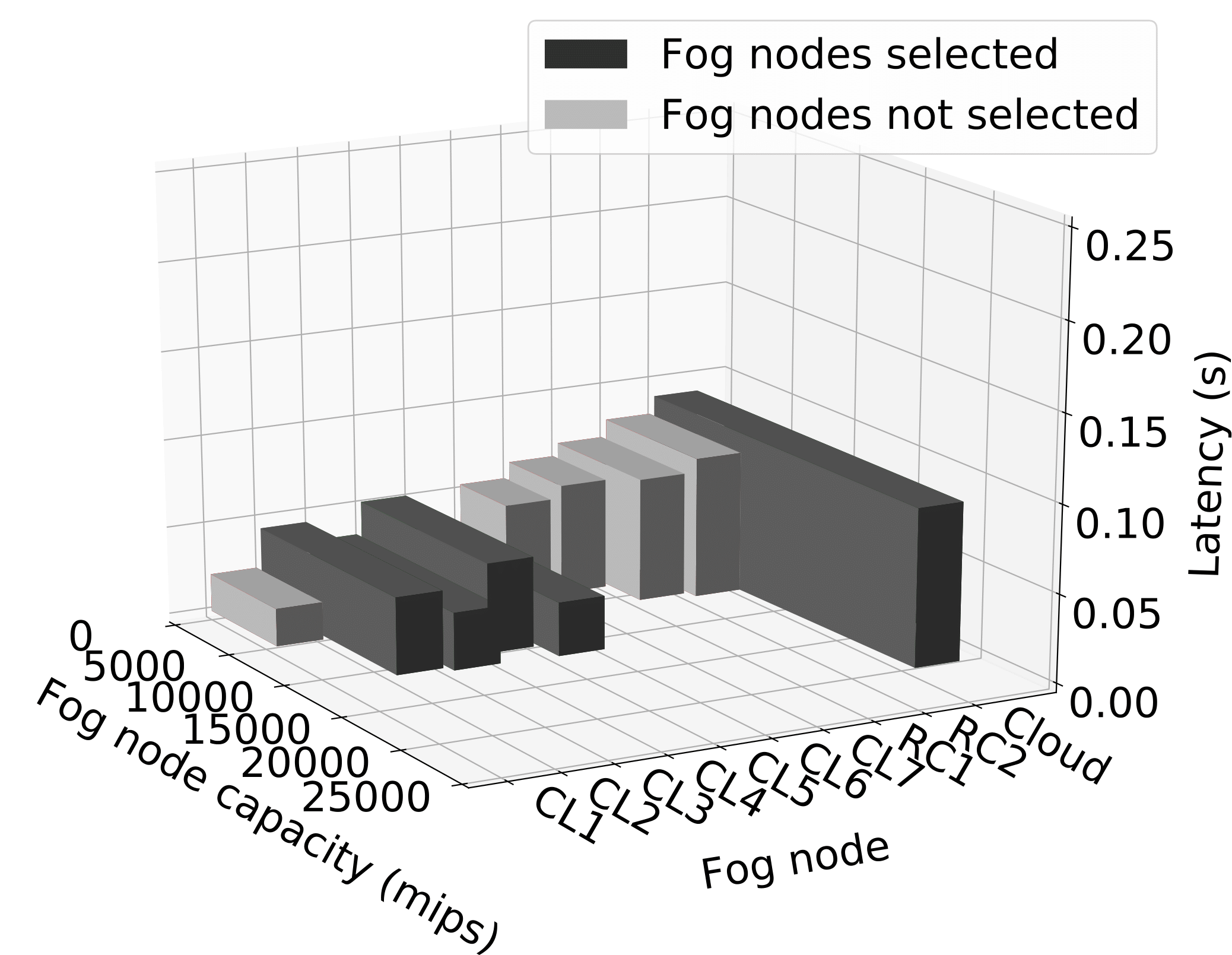}}
    \caption{Medium traffic intensity.}
    \label{fig:region3}
  \end{minipage}

  \bigskip

  \begin{minipage}{.3\linewidth}
    \centering
    \subcaptionbox{Multimedia requests and selected nodes.\label{fig:request4}}
      {\includegraphics[width=\linewidth]{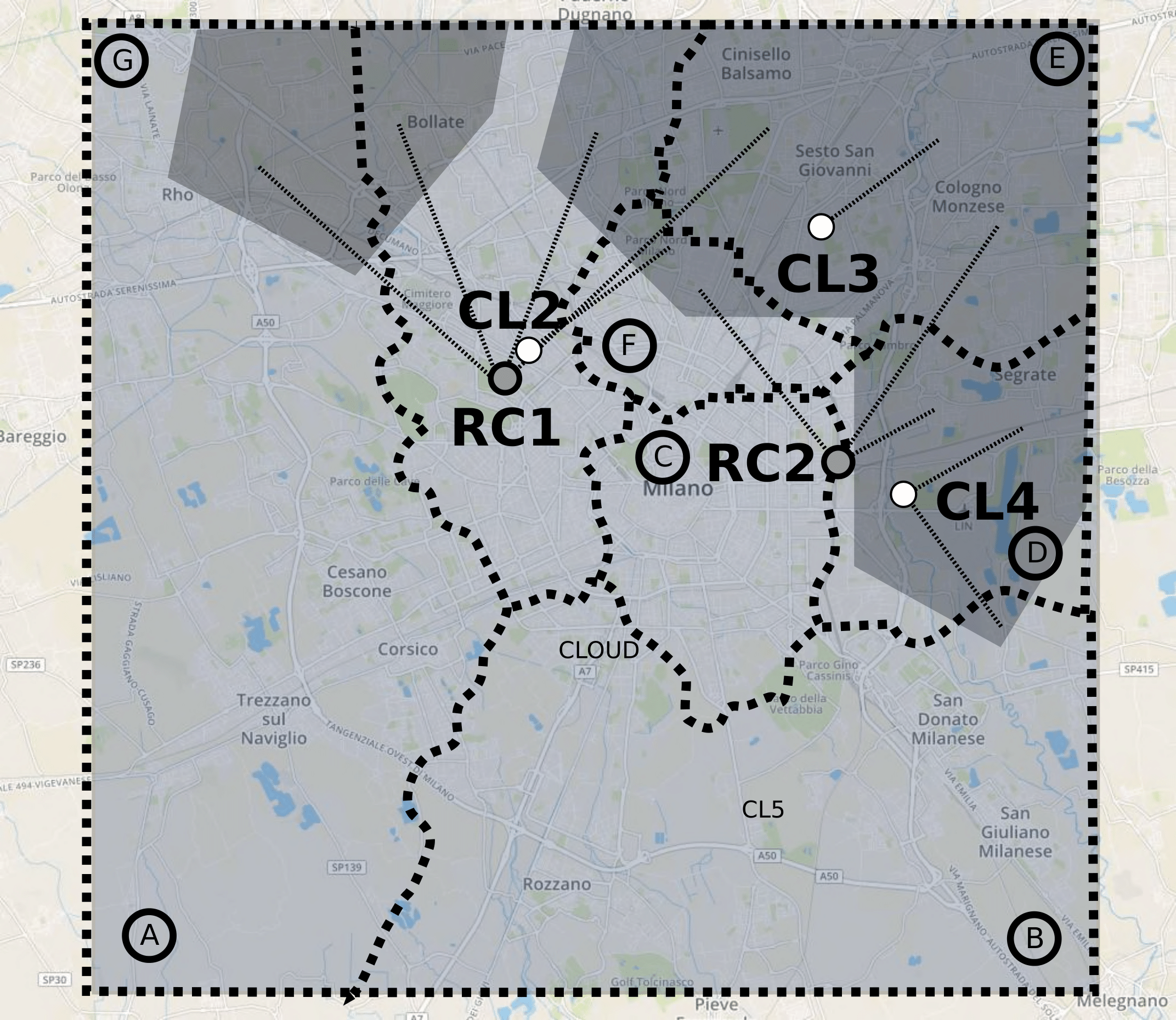}}
    \subcaptionbox{Nodes' hardware capacity and latency.
    \label{fig:fog_nodes_capacity_latency4}}
      {\includegraphics[width=\linewidth]{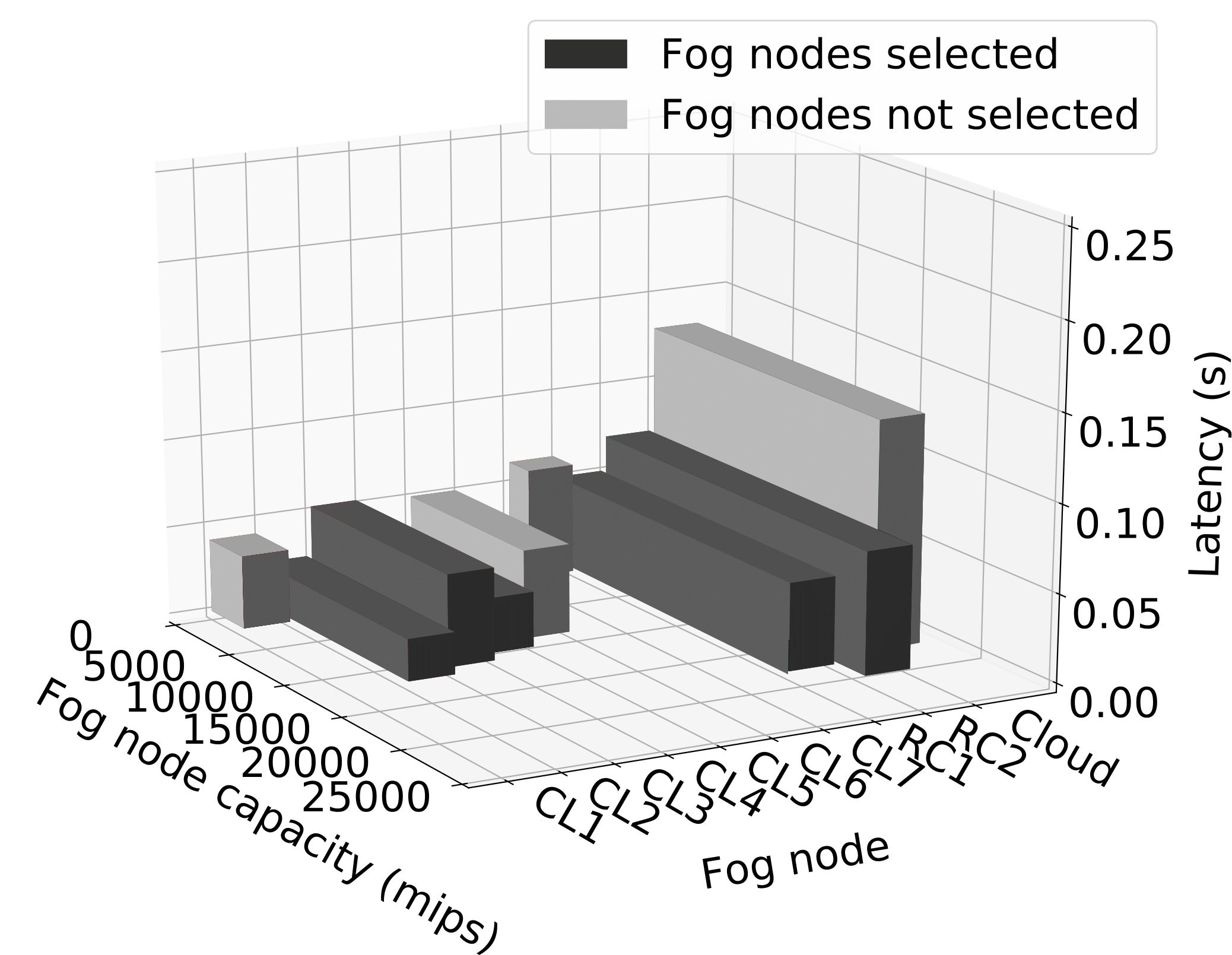}}
    \caption{Medium traffic intensity.}
    \label{fig:region4}
  \end{minipage}
  \quad
  \begin{minipage}{.3\linewidth}
    \centering
    \subcaptionbox{Multimedia requests and selected nodes.\label{fig:request5}}
      {\includegraphics[width=\linewidth]{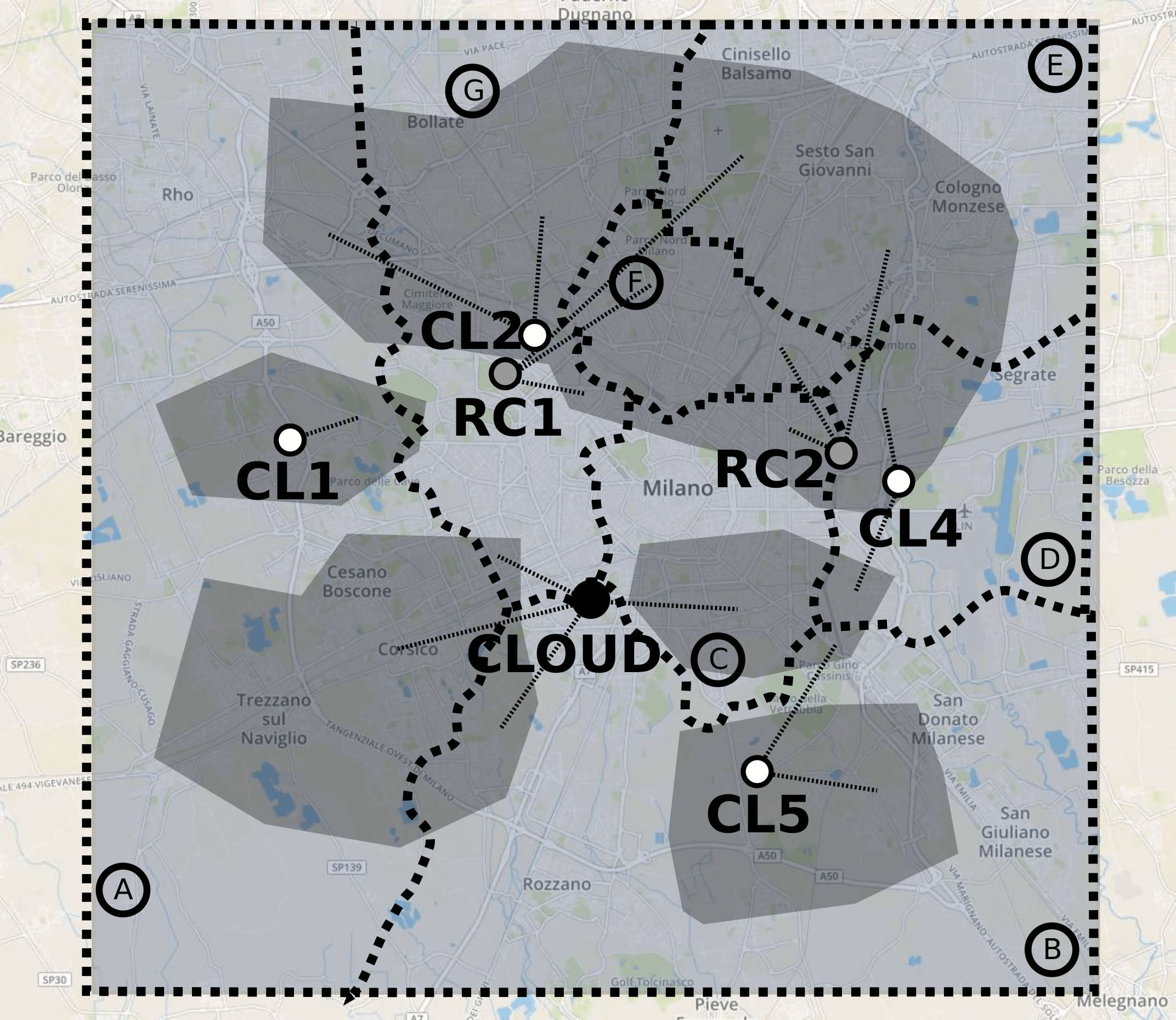}}
    \subcaptionbox{Nodes' hardware capacity and latency.
    \label{fig:fog_nodes_capacity_latency5}}
      {\includegraphics[width=\linewidth]{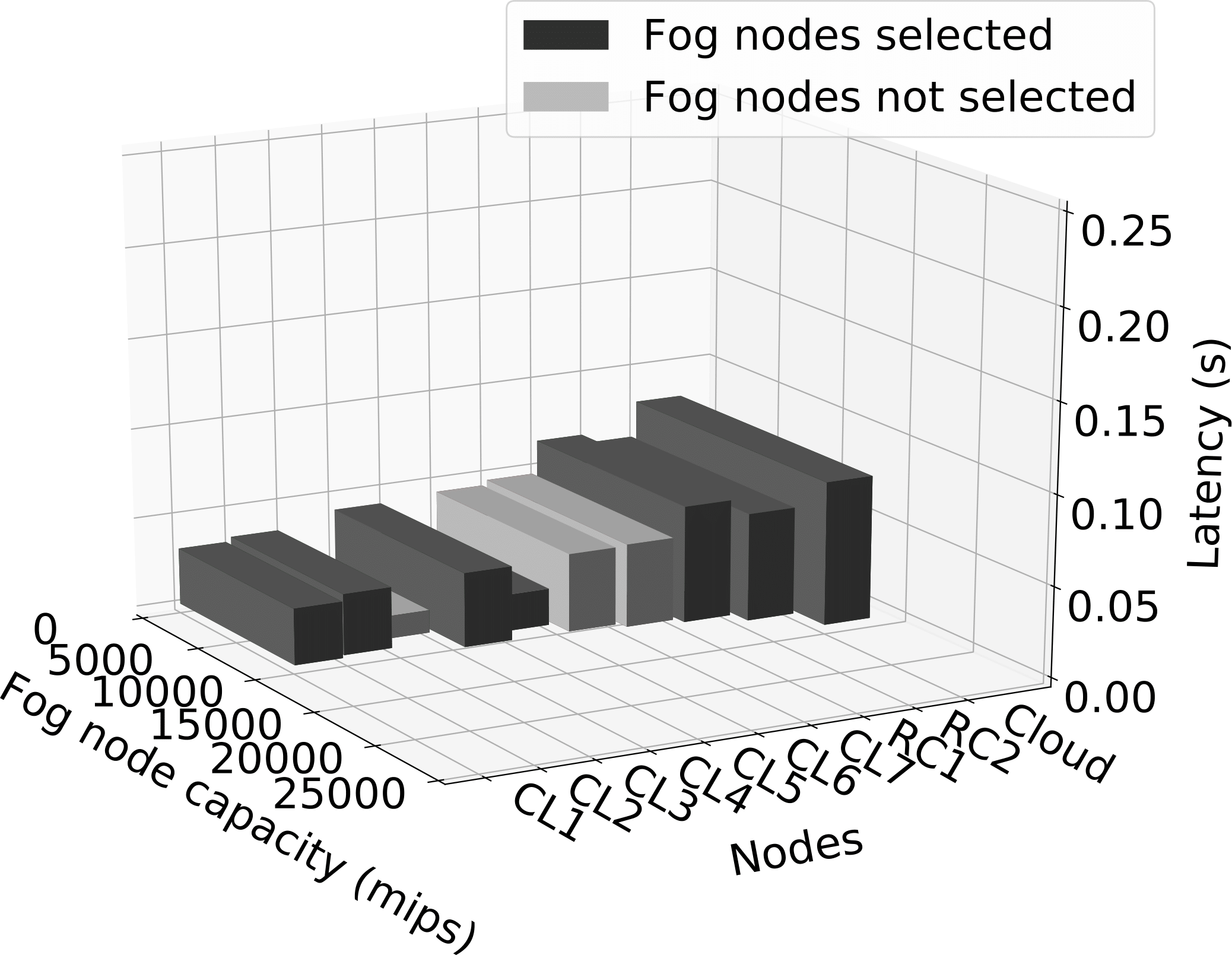}}
    \caption{High traffic intensity.}
    \label{fig:region5}
  \end{minipage}
  \quad
  \begin{minipage}{.3\linewidth}
    \centering
    \subcaptionbox{Multimedia requests and selected nodes.\label{fig:request6}}
      {\includegraphics[width=\linewidth]{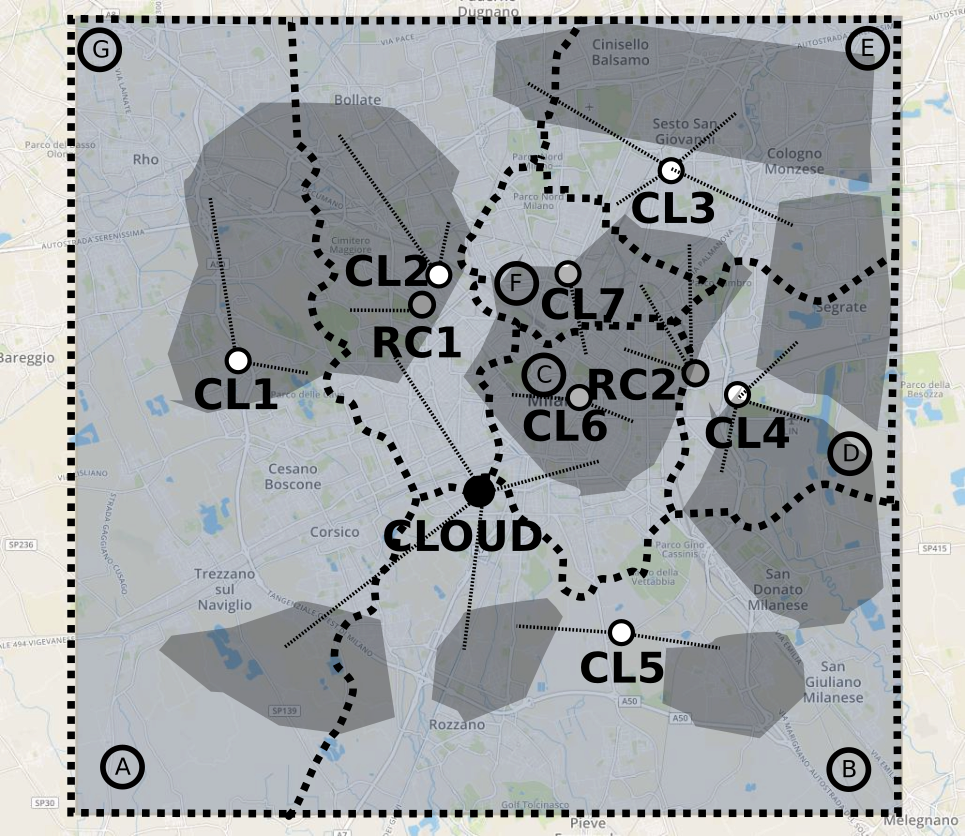}}
    \subcaptionbox{Nodes' hardware capacity and latency.
    \label{fig:fog_nodes_capacity_latency6}}
      {\includegraphics[width=\linewidth]{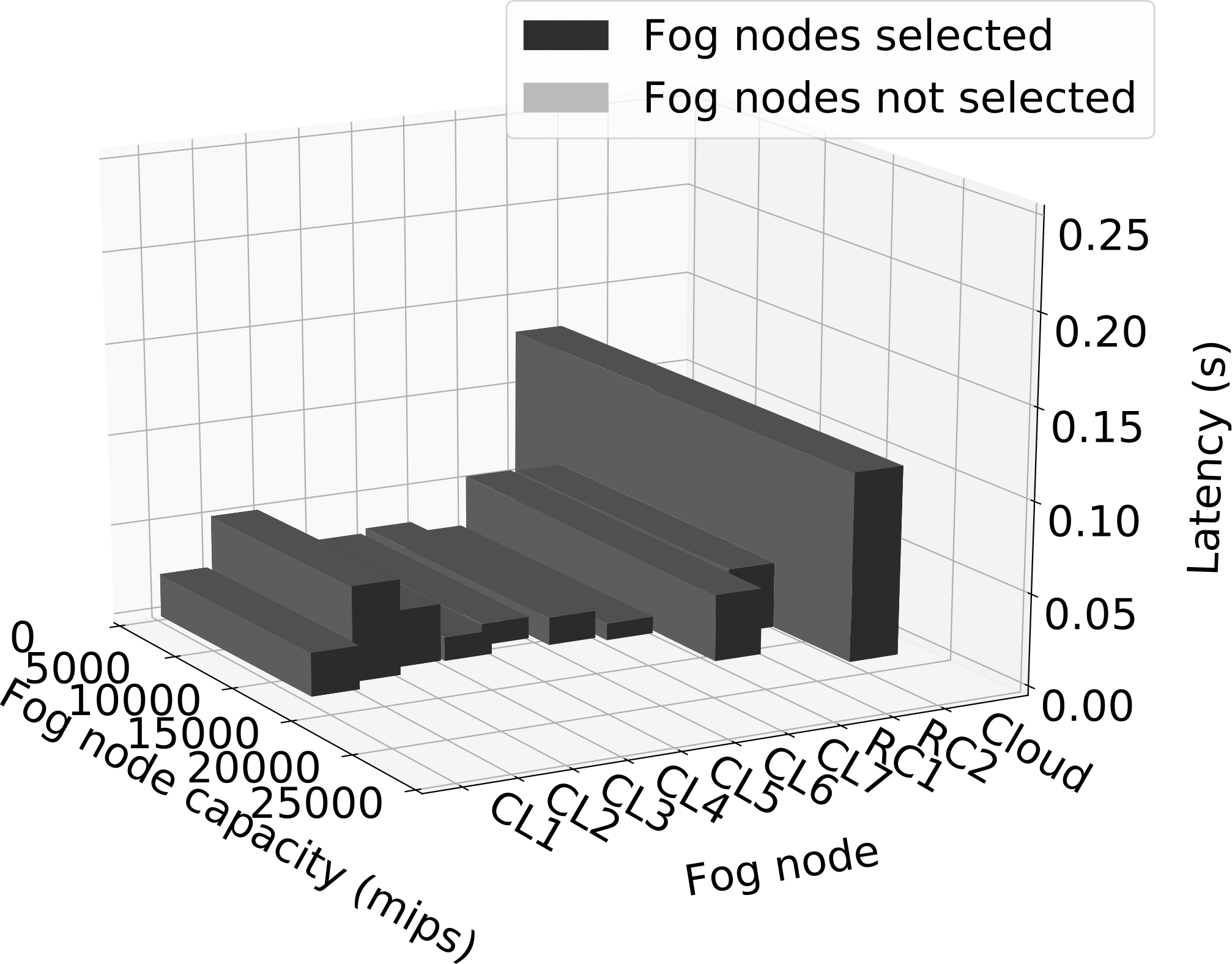}}
    \caption{High traffic intensity.}
    \label{fig:region6}
  \end{minipage}
\end{figure*}

Figures~\ref{fig:region1}-\ref{fig:region6} show the instantaneous multimedia service requests fragmentation measured in the Milan region for each snapshot.
Each one of these figures presents two plots. 
Plots labeled ``(a)'' relate the snapshot of the multimedia service requests in the scenario as well as the nodes selected to receive the multimedia services.
In these plots, every gray spot corresponds to one demand $d_{r}\;\in\;D_{r}$, where $r\;\in\;R$, for multimedia services that must be provided.
The circles represent the nodes enabled to the placement of multimedia services.
Plots ``(b)'' show the nodes' hardware capacity (x-axis) and latency (y-axis) at that moment.
The resulting figures give a rough, yet intuitive, idea of the fog nodes selected in different possibilities.
Additionally, the maximum acceptable delay in delivering multimedia services is less than 0.1 seconds~\cite{liotou2015quality} and the nodes' storage capacity and latency can fluctuate over time.

Figure~\ref{fig:request1} illustrates the first snapshot~(Dec/22/2013 at 05:40).
It is associated with the low traffic intensity group that occurs during dawn on weekends.
Based on this, the nodes \textbf{CL1}, \textbf{CL3}, and \textbf{CL5} are selected.
Figure~\ref{fig:fog_nodes_capacity_latency1} depicts that these Fog nodes have an appropriate hardware capacity and the lowest latency to meet this demand.
Also, they are positioned geographically close to these regions, reducing the latency and may enhancing user experience. 

Figure~\ref{fig:request2} illustrates the second snapshot~(Nov/5/2013 at 10:20). 
It is associated with the medium traffic intensity group that occurs at certain times in the morning.
In this case, all the fog nodes have a maximum acceptable delay in delivering multimedia services, i.e., less than 0.1 seconds.
Also, all cloudlet nodes are low on hardware capacity~(they may be running concurrent services, for example).
Therefore, the nodes \textbf{CLOUD}, \textbf{RC1}, \textbf{and} \textbf{RC2} are selected.
Figure~\ref{fig:fog_nodes_capacity_latency2} illustrates their characteristics at the moment analyzed.

Figure~\ref{fig:request3} illustrates the third snapshot~(Nov/29/2013 at 11:30).
It is associated with the medium traffic intensity group that occurs during the mornings on the weekends.
Again, all nodes have a maximum acceptable delay in delivering multimedia services.
As shown in Figure~\ref{fig:fog_nodes_capacity_latency3}, only the Cloud, and some cloudlet nodes have the hardware capacity to meet this demand. 
The nodes \textbf{CLOUD}, \textbf{CL2}, \textbf{CL3}, \textbf{CL4}, and \textbf{CL5} are selected.

Figure~\ref{fig:request4} illustrates the fourth snapshot~(Nov/10/2013 at 5:00).
It exhibits a medium traffic intensity during dawn on weekends.
Based on Figure~\ref{fig:fog_nodes_capacity_latency4}, the Cloud node CL has high latency ($\geq$ 0.1 s) as CL1, CL5, and CL6 nodes have low hardware capacity to meet such demand.
Therefore, based on demands' geographic location, nodes' hardware capacity, and latency, the nodes \textbf{RC1}, \textbf{RC2}, \textbf{CL2}, \textbf{CL3}, and \textbf{CL4} are selected.

Figure~\ref{fig:request5} illustrates the fifth snapshot~(Dec/21/2013 at 13:40).
It is associated with the high traffic intensity group that occurs during the working hours of weekdays.
Figure~\ref{fig:fog_nodes_capacity_latency5} illustrates that all nodes have a maximum acceptable delay and reasonable hardware capacity.
Again, based on demands' geographic location, nodes' hardware capacity, and latency, the nodes \textbf{CLOUD}, \textbf{RC1}, \textbf{RC2}, \textbf{CL1}, \textbf{CL2}, \textbf{CL4}, and \textbf{CL5} are selected.

Finally, Figure~\ref{fig:request6} illustrates the sixth snapshot~(Dec /21/2013 at 14:30).
Once again, it is associated with the high traffic intensity group that occurs during working hours on Fridays.
Figure~\ref{fig:fog_nodes_capacity_latency6} illustrates that all nodes have a maximum acceptable delay ($\leq$ 0.1 s), but low hardware capacity.
In this special case, all nodes are selected to meet as much of this demand as possible.
Thereby, some regions will not be served, or some users maybe have their video-rate adapted, delivery with poor \gls{qoe} due to the low nodes' hardware capacity.

Considering everything, it is possible to infer that the nodes' storage capacity and geographical location, number of nodes able to process tasks, amount demand for multimedia services, and network latency are all paramount factors to decide which nodes can deploy multimedia services.
Also, selecting these nodes closer to the user using a distributed strategy may reduce the bandwidth consumption, resulting in lower costs and improving network efficiency, guaranteeing that most users who depend on these nodes are served, and improving the network deployed to mitigate provider costs.

\subsection{Performance evaluation of the prediction models}
\label{sec:perfomance-prediction-models}

\gls{arima-pred} generates forecasts through the information contained in the chronological series itself, through the ideal adjustments of its parameters.
Determining them is a non-trivial task~\cite{box2015time}.
However, the $auto\_arima()$ function in Python automatically finds these ideal values.
In general, $\;p\;+\;q\;<=\;2$~\cite{zhang2003time}.
Table~\ref{table:arima-parametros} shows the four best combinations of these parameters generated in relation to the \gls{mae} and \gls{rmse} metrics.

\begingroup
\renewcommand\arraystretch{1}
\begin{table}[H]
\caption{Parameters ($p$,$q$,$d$).}
\centering
\scalebox{1}{
\begin{tabular}{@{}ccc@{}}
\toprule
\textbf{(p,q,d)} & \textbf{MAE} & \textbf{RMSE}\\ 
\midrule 
\textbf{(1,1,1)} & 1.092 & 3.645\\ 
\textbf{(1,1,0)} & 1.108 & 3.985\\ 
\textbf{(0,1,0)} & 1.175 & 4.732\\ 
\textbf{(1,0,0)} & 1.190 & 4.355\\ 
\bottomrule
\end{tabular}}
\label{table:arima-parametros}
\end{table}
\endgroup

\gls{lstm-pred}, based on \gls{lstm} networks, processes data passing on information as it propagates forward.
The training set technique was sliding window, commonly adopted in~\cite{banos2014window}.
The sliding window size is $\;L\;\;=\;144$, including the current timestamp.
This value is based on the daily frequency observed in the series.

\begin{figure}[]
    \centering
        \includegraphics[width=1\linewidth]{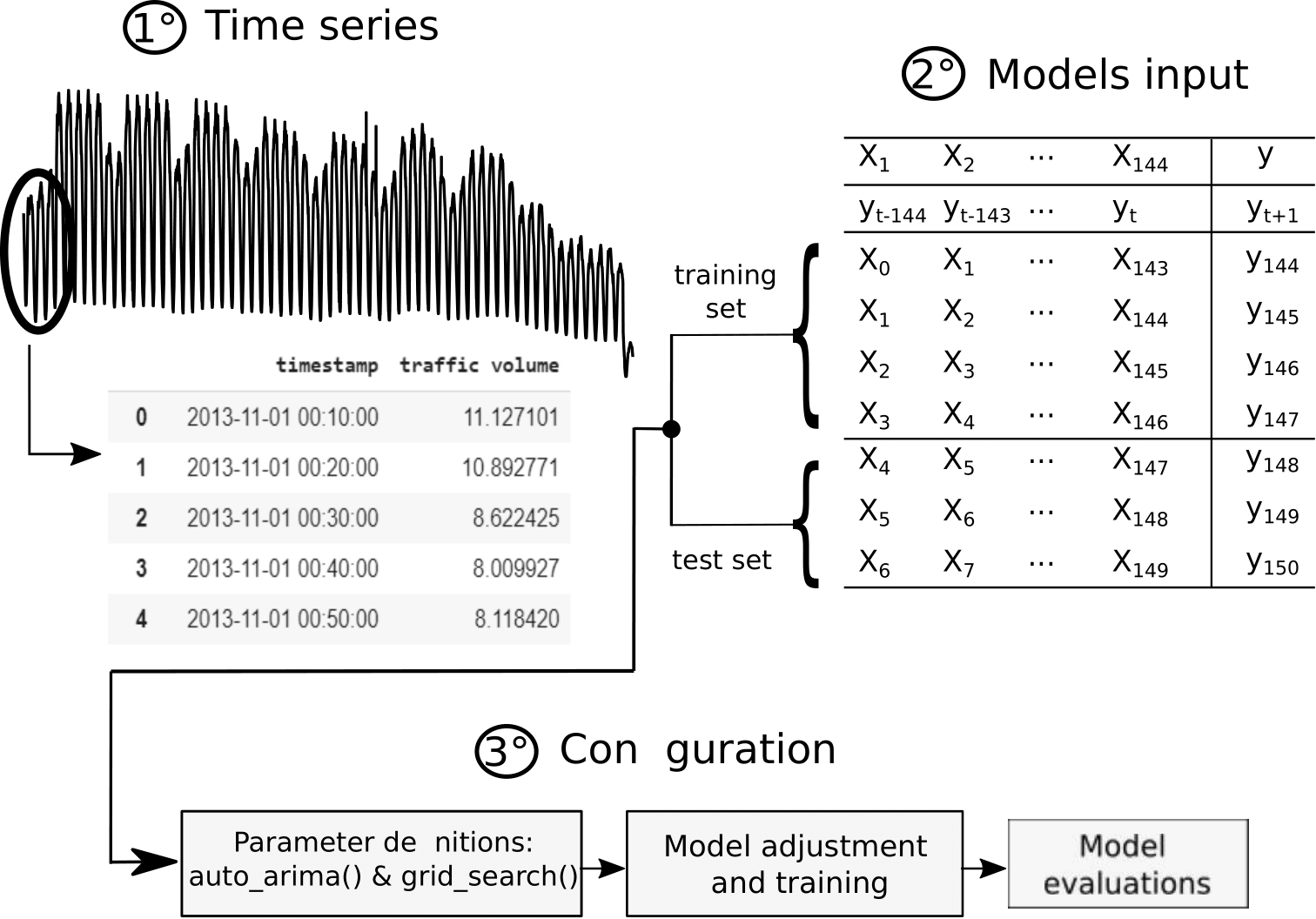}
    \caption{Overall modeling process.}
\label{fig:overall-modeling-process}
\end{figure}

An exhaustive search technique called GridSearchCV is used, provided by the library \textit{sklearn} to find the most suitable hyperparameters.
This technique aims to test all possible combinations of the hyperparameters that were passed to them to find the most suitable configuration for the model in question.
Table~\ref{table:otmz-lstm} shows the values of the estimated hyperparameters that optimize the model.
The first column refers to the hyperparameters.
The second column refers to the estimated values based on~\cite{8667446, prettenhofer2014gradient}.
The third column refers to the values that optimize the model.
The average values for \gls{mae} and \gls{rmse} considering the hyperparameters that optimize the model are $0.97$ and $1.35$, respectively.
Figure~\ref{fig:overall-modeling-process} describes the overall modeling process.

\begingroup
\renewcommand\arraystretch{1}
\begin{table}[H]
\caption{Hyperparameters optimization.}
\centering
\scalebox{.6}{
\begin{tabular}{@{}ccc@{}}
\toprule
\textbf{Hyperparameters} & \multicolumn{1}{c}{\thead{\textbf{Estimated}\\\textbf{values}}} & \multicolumn{1}{c}{\thead{\textbf{Optimized} \\ \textbf{model}}}\\
\midrule 
\textbf{Epochs} & [1500, 1700, 1800] & 1800\\
\textbf{learning rate} & [0.001,0.01,0.1,0.0001] & 0.01\\ 
\textbf{Optimizer} & [Nadam, Adam, RMSProp] & adam\\  
\textbf{Loss function} & [logcosh, mae, mse, hinge, squared\_hinge] &  mae\\
\textbf{Activation function} & [relu, linear, sigmoid, hard\_sigmoid, tanh] & sigmoid\\ 
\textbf{Number of hidden layers} & [1,2,3] & 2\\
\textbf{Dimension of hidden layer} & [200, 400, 600] & 200\\
\midrule
\textbf{Sliding window} & \multicolumn{2}{c}{144}\\
\bottomrule
\end{tabular}}
\label{table:otmz-lstm}
\end{table}
\endgroup

The traffic from the first 40 days is selected as the training data for both models, and the last 22 days' traffic is set to be test data.
The \gls{mae} and \gls{rmse} are adopted as the evaluation metrics and defined as 

\begin{align}
 MAE = (\frac{1}{N})\sum_{i=1}^{N}\left | \hat{y}_{i}-y_{i} \right | \\
 RMSE = \sqrt{\Sigma_{i=1}^{N}{\Big(\frac{\hat{y}_{i}-y_{i}}{N}\Big)^2}}
\end{align}

where $N$ is the sample size, $\hat{y}$ is the prediction and $y_{i}$ the true value.

\begin{figure}
    \centering
        \includegraphics[width=.9\linewidth]{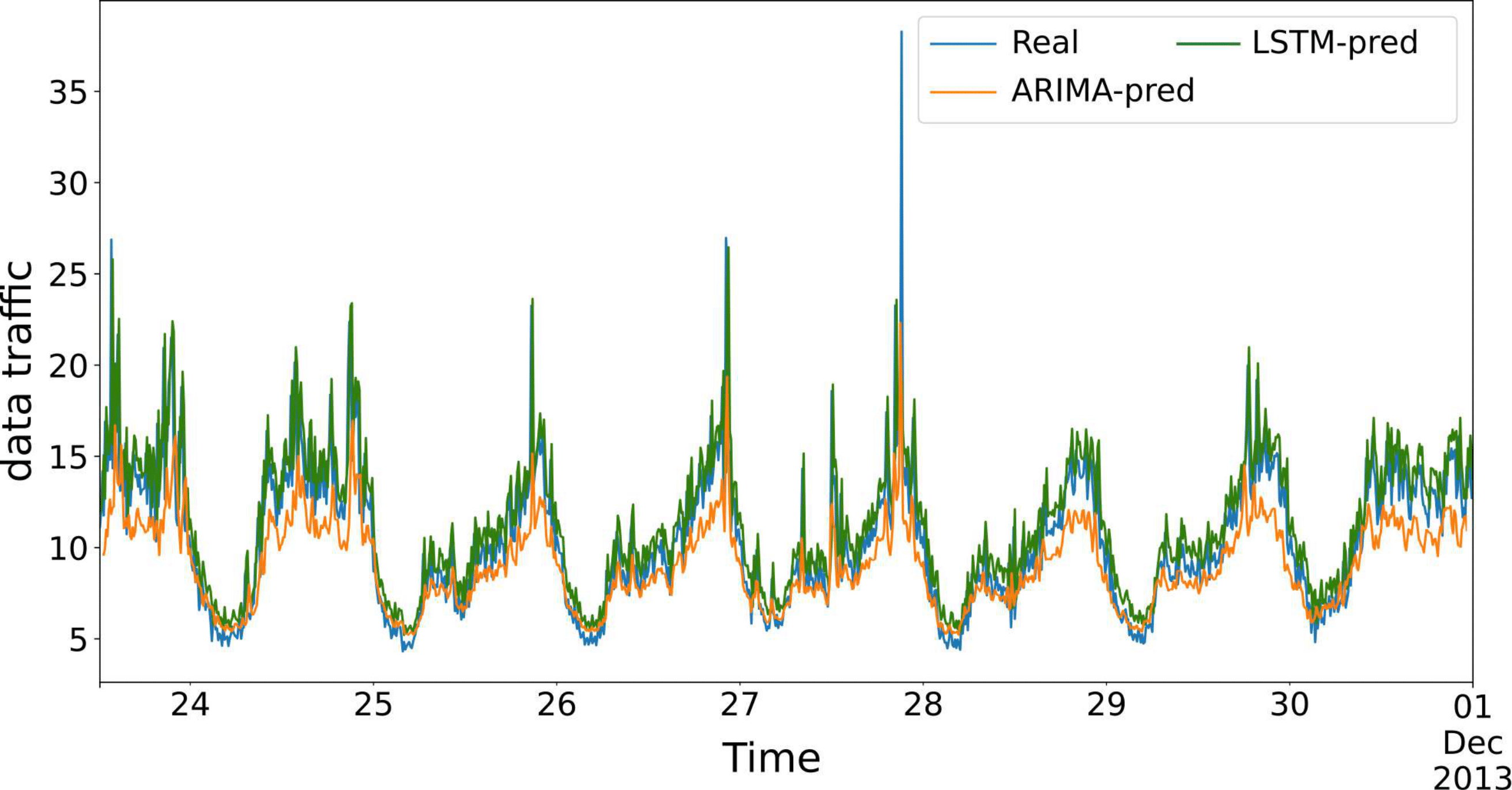}
    \caption{Prediction results considering \gls{arima-pred} and \gls{lstm-pred} models.}
\label{fig:lstm-arima}
\end{figure}

Figure~\ref{fig:lstm-arima} shows the comparisons of traffic data predicted considering a one-month forecast of traffic data.
It can be seen that the prediction results well match the trend of real data traffic even when the traffic becomes unstable due to regular working schedules and urban mobility.

As is possible to notice, the peaks of both in and out traffic can be effectively captured and predicted better by \gls{lstm-pred}.
Due to the irregularity of the time series, the result of the \gls{arima-pred} is generally inferior to those obtained by other methods, such as those based on \gls{rnn}.
On the one hand, models based on \gls{arima} show better results when the series is relatively long and well behaved.
Also, they depend on a greater knowledge of the components of the time series\cite{prettenhofer2014gradient}.
On the other hand, models based on \gls{rnn}s can adapt better to learn the temporal dependencies of the context without much prior knowledge of the series components.
Table~\ref{table:experimental-result} shows the comparison of \gls{mae} and \gls{rmse} considering both models.

\begingroup
\renewcommand\arraystretch{1}
\begin{table}[H]
\caption{Experimental result}
\centering
\scalebox{1}{
\begin{tabular}{@{}ccc@{}}
\toprule
\textbf{} & \textbf{MAE} & \textbf{RMSE}\\ 
\midrule 
\textbf{ARIMA} & 1.092 & 3.645\\ 
\textbf{LSTM} & 0.97 & 1.35\\ 
\bottomrule
\end{tabular}}
\label{table:experimental-result}
\end{table}
\endgroup

Therefore, according to the analysis and the results presented, the prediction model based on the deep learning method (\gls{lstm-pred}) was better than the traditional algorithm model (\gls{arima-pred}).
Thus, \gls{lstm-pred} is utilized in the Model Deployment stage.

\subsection{Network traffic prediction}
\label{sec:traffic-predicted}
The performance evaluation is carried out considering one month of the predicted network traffic in Milan - Italy.
The results are analyzed and compared in terms of (i) content delivery and packet delivery rate; (ii) average latency; and (iii) network usage in terms of (a) total volume of data transmitted during migration and (b) link usage. 
The results are presented with a 95\% confidence interval and compared with four strategies:
\begin{enumerate}
\item \textbf{\gls{da}~\cite{kharel2018multimedia}:} The task for serving the client’s media request is carried out in a hierarchical manner.
\item \textbf{\gls{qoeap}~\cite{mahmud2019quality}:} A linearly optimized mapping of applications and Fog instances that maximizes \gls{qoe}.
\item \textbf{\gls{smart-fl}:} The multimedia services are positioned according to the \textbf{SMART-FL} algorithm.
\item \textbf{\gls{tiptop}:} The multimedia services are positioned according to the \textbf {SMART-FL} algorithm aware of the predicted traffic volume.
\end{enumerate}

\gls{da} strategy starts the search for resources by selecting tiers closest to the user, followed by the others in the hierarchy.
If there are not enough resources in the current tier, the above tier is considered.
\gls{qoeap} presents a centralized service placement approach taking into account both \gls{qoe} and node resources.
\gls{smart-fl} and \gls{tiptop} are solutions proposed by this work.

\begin{figure}[H]
    \centering
    \includegraphics[width=0.7\linewidth]{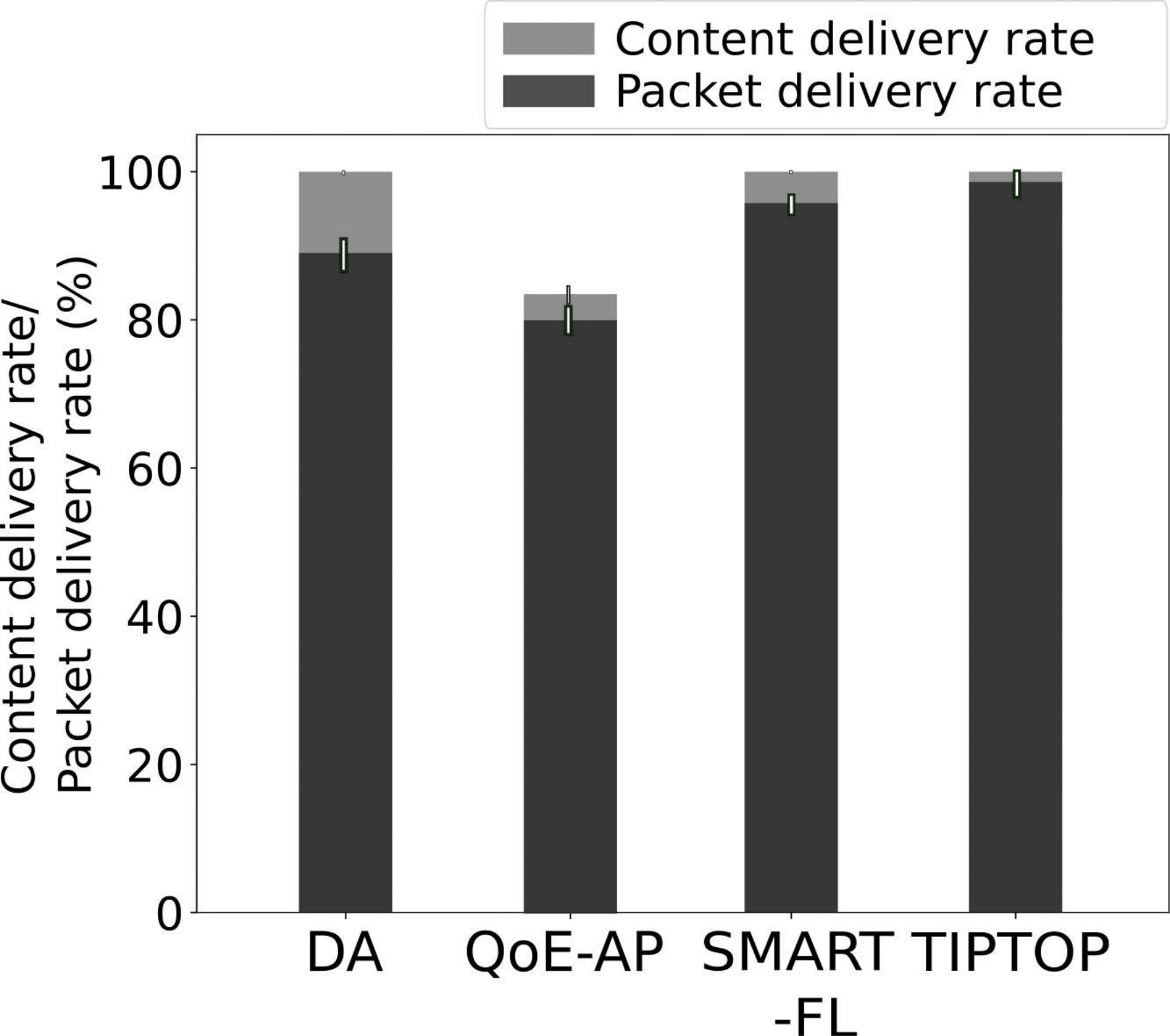}
    \caption{Content delivery and packet delivery rate.}
\label{fig:content_packet}
\end{figure}

Figure~\ref{fig:content_packet} shows the content and packet delivery rate for the four placement strategies considering only requests for multimedia services.

As is possible to notice, \gls{da} strategy presents $\approx\;100\%$ of content delivery rate and $ \approx \; 90 \% $ of packet delivery rate.
This strategy searches for resources by selecting tiers closest to the end-users.
When nodes are selected close to the end-users, the services are offered with high speed and connections with more reliable links, increasing the packet delivery rate, which is not seen here.
In fact, this approach selects the most appropriate tiers, not nodes.

\gls{qoeap} strategy presents $\approx\;83\%$ of content delivery rate and $\approx\;80\%$ of packet delivery rate.
For the same reasons given previously, the content and packet delivery rate are low because the services are offered by nodes not close to the end-users.
This factor, among others, increases the delay and delivers such services with low \gls{qoe}.

In general, both select adequate nodes to provide multimedia services.
However, they do not exploit the benefits of positioning services in different tiers and close to the end-users.
This can increase even more the content and packet delivery rate.
The storage and computation of the services closer to the end-users are performed by the \gls{smart-fl} and \gls{tiptop}.
Both present themselves advantages superior to those discussed so far.

On the one hand, \gls{smart-fl} presents $\approx$ 100 \% of content delivery rate and $\approx$ 96 \% of packet delivery rate.
On the other hand, \gls{tiptop} presents $ \ approx $ 100 \% of content delivery rate and $\approx$ 98 \% of packet delivery rate, which is superior to all strategies.
Both distribute the services across the network and close to users.
The multimedia service placement by \gls{tiptop} is even more adequate due to the set of nodes allocated in $\Gamma$.

\begin{figure}[H]
    \centering
    \includegraphics[width=0.7\linewidth]{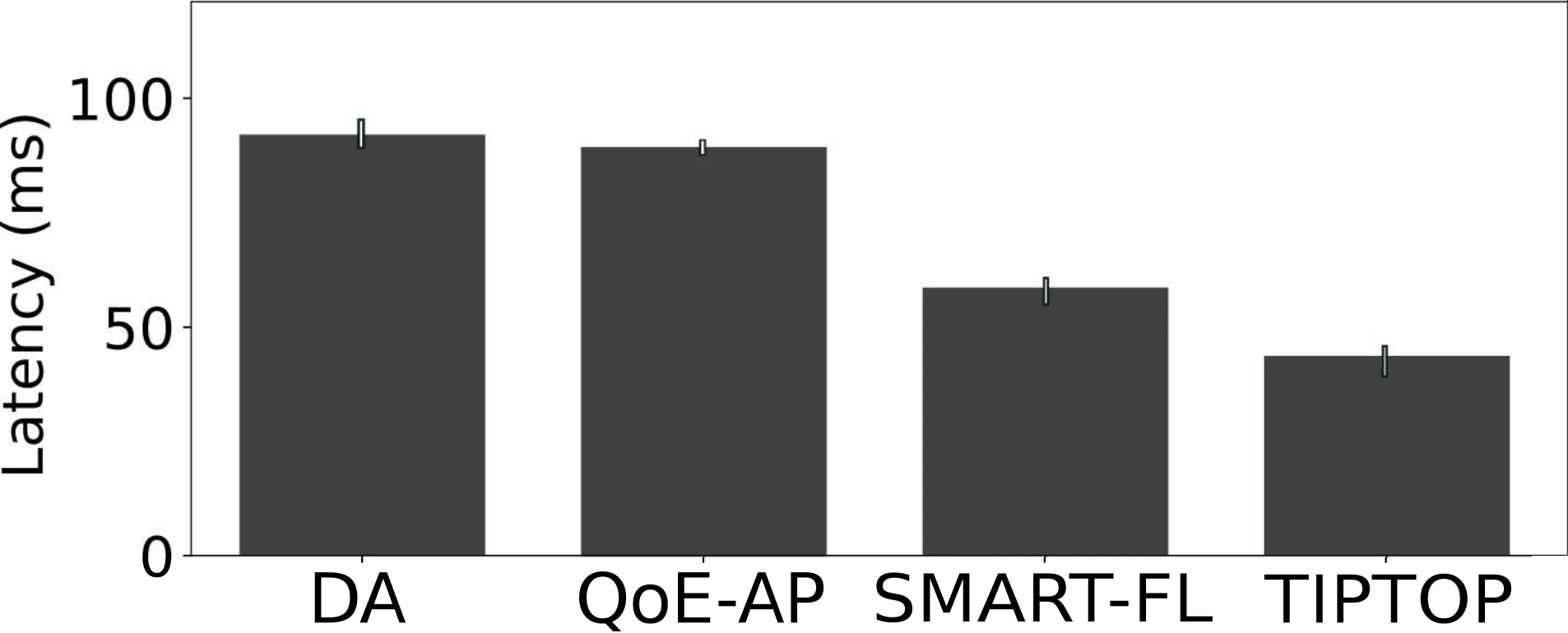}
    \caption{Latency.}
\label{fig:latency}
\end{figure}

Figure~\ref{fig:latency} shows the latency for the four placement strategies considering only requests for multimedia services.

All strategies have an average latency below the maximum acceptable~\cite{liotou2015quality}.
\gls{qoeap} offers services with average latency similar to \gls{da}.
In contrast, \gls{da} presents content and packet delivery rate most appropriate than \gls{qoeap}.

In general, both present average latency adequate to provide multimedia services.
However, this is not sufficient to provide such content that also includes, among others, constant and continuous flow of packets delivery rate, not seen in both.

\gls{smart-fl} and \gls{tiptop} strategies provide constant and continuous flow by selecting nodes distributed throughout the network and close to the end-users.
This decreases the number of hops and reduces the latency to deliver these services through connections with more reliable links.

In comparison, the \gls{tiptop} strategy offers the lowest average latency among all strategies.
The reason is that the nodes in $\Gamma$ are in tiers closer to the users and consequently have lower latency than the nodes in $A_{t(n + 1)}$.

\begin{figure}[H]
    \centering
    \includegraphics[width=0.7\linewidth]{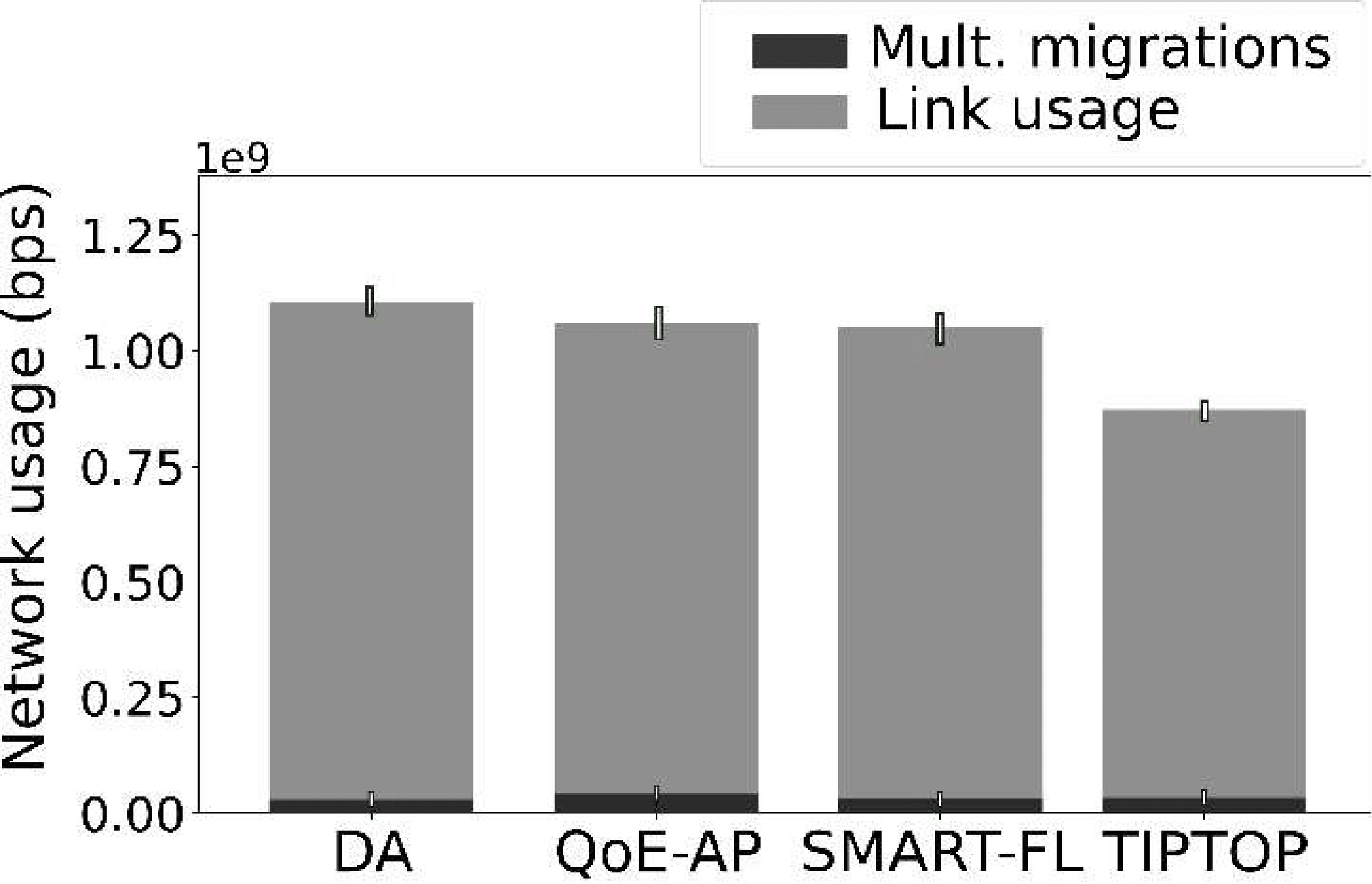}
    \caption{Network usage in terms of (a) total volume of data transmitted during migration and (b) link usage.}
\label{fig:net-work-usage}
\end{figure}

Figure~\ref{fig:net-work-usage} shows the network usage in terms of (a) total volume of data transmitted during migration and (b) link usage, considering only requests for multimedia services.

\gls{da} strategy presents a higher network usage than all of them.
This is due to the non-distribution of multimedia services at the network edge, which reduces the overall network usage because fewer channels for data transmission are utilized.
Also, a cost of $\approx \; 29391592.17$ \gls{bps} is added due to multimedia services migrations to the most appropriate tiers due to variations in network resources and nodes' hardware capacity.

\gls{qoeap} strategy presents an adequate average network usage.
However, as discussed, it presents the lowest content and packet delivery rate between all strategies.
Also, it presents a similar cost for multimedia service migrations.
It seems that \gls{qoeap} strategy evaluated in a strongly connected and fully Cloud-Fog-enabled scenario presents underestimated results.

As mention before, the benefits of distributing services across the network are not explored for both strategies, which can further reduce network usage since fewer channels for data transmission are used.
This and others benefits are provided by \gls{smart-fl} and \gls{tiptop} strategies. 

\gls{smart-fl} strategy presents a network usage similar to \gls{da} and \gls{qoeap}.
However, it presents a better content rate, packet delivery rate, and latency.
Likewise, a cost of $\approx \; 31267651.24$ \gls{bps} is added.
This cost occurs because of the same reasons given previously.
It is a minimal cost about the advantages presented.
This strategy has lower than maximum acceptable latency, adequate content and packet delivery rate, and relatively low network usage.

\gls{tiptop} strategy presents even more satisfactory results.
The network usage is lower than all strategies presented.
Likewise, a cost of $\approx \; 34176269.96$ \gls{bps} is added.
This cost occurs because of the same reasons given previously.
As mentioned before, it is a minimal cost about the advantages presented.
Nodes reserved in $\Gamma$ have advantages over nodes in $A_{t(n + 1)}$.
Therefore, of all the strategies, this is the one that most benefits from the available resources and environment.

\section{Conclusions and future work}
\label{sec:conclusion}

The constant communication technology's advancements and the great availability of streaming video services require new methods to ensure the quality for end-users.
To improve on this issue, this work first presented a process to design/create a hierarchical multi-tier Cloud-to-Fog network for the distribution of multimedia services.
Moreover, this work also proposed an algorithm that selects the minimum number of nodes able to deliver multimedia services with low latency in multi-tier Fog nodes architecture.
By the way, the traffic forecast improves the provision of these services.
The performance assessment was carried out using two months of real-world mobile network traffic data in Milan, Italy. 

Instead of concentrating data and computation in a small number of large Clouds, we consider that either portions or all of the Cloud services must migrate to fog nodes located closer to the user, meeting their needs regarding latency.
This is one of the factors that form the primary motivation of this paper.

Keeping this analysis closer to the data source, especially for latency-sensitive services where every millisecond is essential, in addition to the advantages presented in this work, may improve the user experience and reduce overhead in the Cloud as a whole~\cite{bonomi2012fog}.
Moreover, reducing the demand on the Cloud allows turning off servers in the data center to save energy.
As has been noted, the proposed solution can be used in a real-world network to cope with future challenges in providing seamless and, at the same time, high-quality multimedia services in hierarchical multi-tier Fog nodes.

Future work intends to extend the proposed method utilizing new approaches, such as \gls{dbscan}, to find communities and analyze the energy consumption and network usage.
The proposed algorithm for the multimedia services placement is based on an \gls{ilp} model. The \gls{ilp} model is logically based on linear equations.  However, some decision variables, such as \gls{qoe} and \gls{qos}, have a non-linear effect, which are not considered by the \gls{ilp} model. More research could be carried out in this direction.
Also, more dynamic evaluation scenarios will be used to prove the benefits of the \gls{tiptop}.

\section*{Acknowledgments}
The authors acknowledge support from the Brazilian research agency CNPq, CAPES and INCT of the Future Internet for Smart Cities (CNPq 465446 / 2014-0, CAPES 88887.136422 / 2017-00, and FAPESP 2014 / 50937-1).

\section{References}

\bibliographystyle{elsarticle-num}
\bibliography{main}

\end{document}